\documentclass[traditabstract]{aa}
\usepackage{color,graphicx}
\usepackage{txfonts}
\usepackage{natbib}
\newcommand\Lsun{{L$_{\sun}$}}
\newcommand\Msun{{M$_{\sun}$}}

\newcommand\micron{{$\mu$m}}
\newcommand\Tdust{{T_{\mathrm{dust}}}}

\definecolor{orange}{rgb}{1.0,0.6470588235294118,0.0}
\definecolor{blue}{rgb}{0.0,0.0,1.0}
\definecolor{green}{rgb}{0.0,0.50196078431372548,0.0}
\definecolor{red}{rgb}{1.0,0.0,0.0}
\definecolor{magenta}{rgb}{1.0,0.0,1.0}
\definecolor{purple}{rgb}{0.50196078431372548,0.0,0.50196078431372548}

\begin{document}

\title{On the massive young stellar object AFGL~4176}
\subtitle{High-spatial-resolution multi-wavelength observations and modeling}
\author{Paul A. Boley
  \inst{1}
  \and
  Hendrik Linz
  \inst{1}
  \and
  Roy van Boekel
  \inst{1}
  \and
  Jeroen Bouwman
  \inst{1}
  \and
  Thomas Henning
  \inst{1}
  \and
  Andrey M. Sobolev
  \inst{2}
  }

\institute{Max Planck Institute for Astronomy, K\"onigstuhl 17, Heidelberg, Germany \\\email{[boley,linz,boekel,bouwman,henning]@mpia.de}
  \and
  Ural Federal University, Astronomical Observatory, 51 pr. Lenina,
  Ekaterinburg, Russia \\\email{andrey.sobolev@usu.ru}}

\date{\today}

\abstract{Deeply embedded and at distances of several kiloparsecs,
  massive young stellar objects (MYSOs) present numerous challenges
  for observation and study.  In this work, we present
  spatially-resolved observations of one MYSO, AFGL~4176, together
  with survey and literature data, ranging from interferometric
  observations with VLTI/MIDI in the mid-infrared, to single-dish
  Herschel measurements in the far-infrared, and sub-millimeter data
  from APEX.  We consider this spatially-resolved, multi-wavelength
  data set in terms of both radiative transfer and geometric models.
  We find that the observations are well described by one-dimensional
  models overall, but there are also substantial deviations from
  spherical symmetry at scales of tens to hundreds of astronomical
  units, which are revealed by the mid-infrared interferometric
  measurements.  We use a multiple-component, geometric modeling
  approach to explain the mid-infrared emission on scales of tens to
  hundreds of astronomical units, and find the MIDI measurements are
  well described by a model consisting of a one-dimensional Gaussian
  halo and an inclined ($\theta=60\degr$) circumstellar disk extending
  out to several hundred astronomical units along a position angle of
  160\degr.  Finally, we compare our results both with previous models
  of this source, and with those of other MYSOs, and discuss the
  present situation with mid-infrared interferometric observations of
  massive stars.}

\keywords{stars: formation - techniques: interferometric - techniques:
  high angular resolution - radiative
  transfer - stars: individual: AFGL~4176}
\maketitle

\section{Introduction}

The class of objects known as massive young stellar objects (MYSOs)
are thought to represent massive stars at the earliest phases of
evolution.  The inherently large distances (several kpc) and high
extinctions ($A_V \ga 20$~mag) to typical MYSOs complicate attempts to
characterize these objects observationally, and many studies
\citep[e.g.][]{Grave09} rely on fitting the spectral energy
distribution (SED) to untangle the underlying physics involved in the
formation of massive stars.

However, it has been shown that pure SED fitting, i.e. without making
use of any spatial information, is a highly degenerate approach, at
best \citep[e.g.][]{Thamm94,Menshchikov97}.  Different physical scales
and temperature regimes of the circumstellar material can be resolved
using a variety of techniques at different wavelengths, ranging from
weakly-resolved single-telescope measurements, spectroscopic
(spectroastrometric) measurements, and interferometric measurements,
both at near-/mid-infrared and (sub-)millimeter wavelengths.  The
temperatures and regions thus probed range from $\sim10^3-10^4$~K gas
within the dust sublimation radius using spectroastrometry of
molecular gas lines \citep[e.g.][]{Wheelwright10}, $\sim10^3$~K dust
at the inner dust disk radius using near-infrared interferometry
\citep[e.g.][]{Kraus10}, $\sim10^2-10^3$~K dust farther out in the
disk (and possibly into outflow cones) using mid-infrared
interferometry \citep[e.g.][]{Linz09}, and the broad distribution of
cold $\sim10^1-10^2$~K dust using millimeter-wave interferometry
\citep[e.g.][]{Beuther07} and single-dish measurements
\citep[e.g.][]{Beltran06}.

In this paper, we present observations and the results of
multi-wavelength modeling of one particular massive young stellar
object, AFGL~4176 (also known as IRAS~13395-6153, G308.9+0.1).  This
object was identified as a candidate MYSO based on the similarity of
the far-infrared spectrum to that of the Becklin-Neugebauer (BN)
object in Orion \citep{Henning84,Henning90}.  The class of so-called
BN-type objects are luminous ($\ga 10^3$~\Lsun) sources, extremely red
(usually with no optical counterpart), with a broad peak in emission
(flat spectrum in $\nu F_\nu$) between 50 and 200~\micron{} and deep
absorption due to silicates at 10~\micron{}.  They are generally
interpreted as young (possibly still forming) massive stars,
surrounded by a thick dust envelope, despite the unclear nature of the
prototype object Orion BN (see the work by \citet{Tan04}, for
example).

At radio wavelengths, using the Molonglo Observatory Synthesis
Telescope (MOST), \citet{Caswell92} reported a small-diameter source
($\le$0\farcm2) roughly coincident with the IR source AFGL~4176.  The
integrated flux densities are 0.31~Jy at 843~MHz and 0.40~Jy at
1415~MHz, confirming the emission is thermal bremsstrahlung radiation
arising from an \ion{H}{II} region.  Using the Australian Telescope
Compact Array (ATCA), \citet{Phillips98} reported compact continuum
emission at 8.6~GHz spanning about 10\arcsec{}, with a collection of
6.7~GHz methanol masers on the northern edge.  Later,
\citet{Ellingsen05} showed the 8.6~GHz continuum emission extends to
about 1\arcmin.  Finally, the 1.2~mm dust continuum was mapped by
\citet{Beltran06}, and comprises a roughly circular, $\sim
40$\arcsec{} clump at the same location with an estimated mass of
1120~\Msun{} at 5.3~kpc, and a fainter clump (estimated mass
177~\Msun{} at 5.3~kpc) about 1\farcm5 to the east.

In the near infrared, AFGL~4176 is a point source for seeing-limited
observations.  Using speckle interferometry, however,
\citet{Leinert01} found a faint extended halo in the $K$ band, with a
size of $1\farcs6\pm0\farcs4$.  The positions from the Two Micron All
Sky Survey (2MASS) and the methanol masers reported by
\citet{Phillips98} are within 0\farcs7.  \citet{deBuizer03} detected a
large number of shocked H$_2$ emission blobs around AFGL~4176 by means
of narrow-band NIR imaging. However, their spatial distribution does
not reveal a preferred direction for a potential jet, and in general,
the association of AFGL~4176 with a jet or outflow remains unclear.
The source was also observed in the mid- through far- infrared by the
space missions IRAS, MSX, ISO, Spitzer and Herschel.  We note that the
coordinates listed in the IRAS Point Source Catalog appear to be
erroneous when compared with the MSX 12~\micron{} and Spitzer
24~\micron{} images, and adopt the 2MASS position of
$\alpha=13^\mathrm{h}43^\mathrm{m}01\fs70$
$\delta=-62\degr08\arcmin51\farcs23$ (J2000) as the location of the
infrared source.  The ISO spectrum shows deep silicate absorption
features at 9.7 and 18~\micron{}, as well as absorption due to water
ice at 3~\micron{}, CO$_2$ ice at 4.3~\micron{}, and emission lines of
[\ion{Ne}{II}] at 12.8~\micron; [\ion{S}{III}] at 12.0, 18.7 and
33.5~\micron; [\ion{O}{I}] at 145.5~\micron{} and possibly
63.2~\micron; and [\ion{C}{II}] at 157.7~\micron{}.

The distance to AFGL~4176 remains poorly constrained.  Using both the
velocity curve of \citet{Brand93} with updated parameters from
\citet{Levine08}, and the velocity curve of \citet{Reid09}, we find
that the combination of the source's LSR velocity of $\sim
-51$~km~s$^{-1}$ \citep[e.g.][]{Fontani05,deBuizer09} and galactic
coordinates are forbidden\footnote{\label{footnote_distance}We note
  that the tangential solution of these velocity curves at the
  coordinates of AFGL~4176 occurs for $V_{\mathrm{LSR}} \approx
  -49$~km~s$^{-1}$ and corresponds to about 5.3~kpc, which may explain
  the kinematic distances reported by other authors.}. Consequently, a
kinematic distance cannot be determined for AFGL~4176 using these
rotation curves.  Despite this, several authors
\citep[e.g.][]{Beltran06,Fontani05,Grave09} report kinematic distances
of around $5$~kpc.  \citet{Saito01} showed that the spread in
velocities of C$^{18}$O molecular clumps in the vicinity is high
($\sim 30$~km~s$^{-1}$), meaning that kinematic estimates of the
distance to this object should be highly uncertain, even after
improvement of the galactic structure parameters.  To date, other
means of the distance measurements (e.g. maser parallax measurements)
have not been applied to AFGL~4176.  Because of this uncertainty, we
consider the observational data in terms of both ``near'' (3.5~kpc)
and ``far'' (5.3~kpc) distances, corresponding to the edges of the
large C$^{18}$O complex as determined by \citet{Saito01}.

The present paper is similar in scope to previous infrared
interferometric studies of MYSOs by
\citet{Linz09,Follert10,deWit10,Kraus10}.  However, in this work on
AFGL~4176, we go beyond traditional SED modeling techniques and use a
far more comprehensive approach, including not only flux values, but
also spatial information at mid-infrared, far-infrared, and
sub-millimeter wavelengths directly into the fitting process.  Such an
approach helps to break the degeneracies inherent in modeling the SEDs
of deeply-embedded sources, and provides structural information on
scales ranging from tens of AU up to a parsec.  We begin by presenting
and discussing new observations of AFGL~4176, then go to
one-dimensional radiative transfer modeling of the entire data set.
Next, we use a geometric component analysis of the mid-infrared
interferometric data to explore deviations from spherical symmetry,
which we interpret as strong evidence for a circumstellar disk around
this object.  Finally, we compare the results of our study with those
of previous studies of this object, as well as studies of other MYSOs.

\section{Observations and data reduction}

\subsection{Mid-infrared interferometry}
\label{sec_midiobs}

\begin{figure}[b]
  \begin{center}
    \includegraphics[width=85mm]{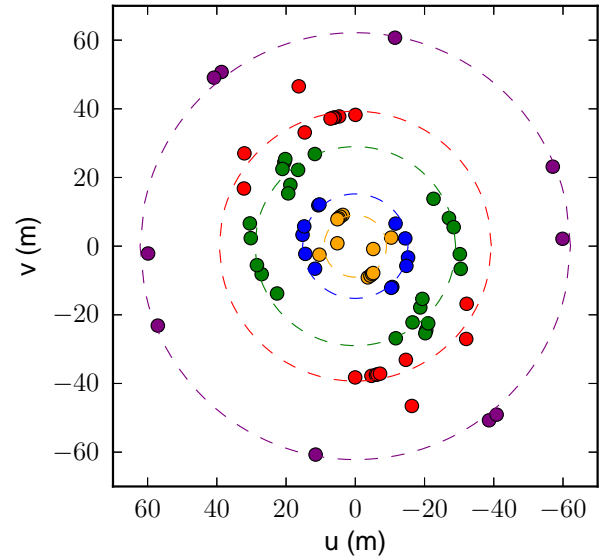}
    \caption{$uv$ coverage of MIDI observations.  North is up, east is
      left.  The positions have been color-coded by projected baseline
      length, and the dashed lines indicate the mean projected
      baselines for each color (\emph{orange} -- 8.1~m; \emph{blue} --
      15.2~m; \emph{green} -- 28.5~m; \emph{red} -- 42.5~m;
      \emph{purple} -- 61.7~m).}
    \label{fig_uvplot}
  \end{center}
\end{figure}

AFGL~4176 was observed with the two-telescope mid-infrared
interferometer MIDI on the Very Large Telescope Interferometer (VLTI)
of the European Southern Observatory on Cerro Paranal in Chile.
Observations were conducted in 2005-2007 and 2011-2012 with both the
8.2~m Unit Telescopes (UTs) and the 1.8~m Auxiliary Telescopes (ATs)
as part of guaranteed-time observations (proposal IDs 074.C-0389,
075.C-0755, 076.C-0757, 077.C-0440, 078.C-0712 and 084.C-1072).  In
total, 37 points in the $uv$ plane were obtained for projected
baselines ranging from 5 to 62 meters and a variety of position
angles, with six of the measurements obtained with the UT telescopes,
while the remaining 31 were obtained with the AT telescopes.  This
represents, by far, the largest collection of long-baseline
mid-infrared interferometric observations presented for a MYSO to
date.  A summary of the observations is shown in
Table~\ref{tab_obslog}, and the $uv$ coverage is shown in
Fig.~\ref{fig_uvplot}.  For clarity, the individual observations have
been color-coded by projected baseline length: purple indicates
measurements with the longest baselines (60-64~m); red (36-49~m) and
green (25-32~m) show those with intermediate baselines; blue (13-16~m)
and orange (5-11~m) mark the shortest baseline measurements.

\begin{table*}
\begin{center}
\caption{Observation log}
\label{tab_obslog}
\begin{tabular}{l l c c c}
\hline \hline
Date/time & Telescopes & Proj. baseline & Position angle & Avg. visibility\\
(UTC) & & (m) & (deg.) & \\
\hline
\textcolor{purple}{2005-03-02~~08:08} & UT3-UT4 & 61.6 & 112.1 & 0.02\\
\textcolor{purple}{2005-06-24~~04:54} & UT3-UT4 & 61.8 & 169.3 & 0.02\\
\textcolor{red}{2005-06-25~~22:50} & UT1-UT2 & 49.3 & 19.3 & 0.07\\
\textcolor{red}{2005-06-26~~02:05} & UT1-UT2 & 41.9 & 49.8 & 0.08\\
\textcolor{red}{2005-06-26~~03:27} & UT1-UT2 & 36.3 & 62.4 & 0.04\\
\textcolor{green}{2005-06-27~~03:31} & UT2-UT3 & 30.3 & 85.6 & 0.08\\
\textcolor{blue}{2006-02-22~~05:46} & E0-G0 & 16.0 & 41.9 & 0.30\\
\textcolor{green}{2006-04-18~~08:01} & D0-G0 & 26.5 & 121.4 & 0.06\\
\textcolor{blue}{2006-04-24~~07:30} & E0-G0 & 13.4 & 119.4 & 0.31\\
\textcolor{blue}{2007-02-11~~06:24} & E0-G0 & 16.0 & 40.5 & 0.23\\
\textcolor{green}{2007-02-12~~06:18} & G0-H0 & 31.9 & 40.0 & 0.13\\
\textcolor{green}{2007-03-15~~07:08} & G0-H0 & 31.2 & 77.7 & 0.11\\
\textcolor{green}{2007-03-16~~09:13} & G0-H0 & 28.3 & 106.8 & 0.07\\
\textcolor{blue}{2007-04-18~~04:55} & E0-G0 & 15.6 & 77.8 & 0.20\\
\textcolor{blue}{2007-05-24~~01:51} & E0-G0 & 15.8 & 68.7 & 0.27\\
\textcolor{green}{2007-06-19~~02:35} & G0-H0 & 28.9 & 101.0 & 0.06\\
\textcolor{blue}{2007-07-03~~01:31} & E0-G0 & 14.6 & 98.9 & 0.28\\
\textcolor{orange}{2011-05-09~~02:15} & B2-C1 & 9.8 & 21.7 & 0.47\\
\textcolor{orange}{2011-05-09~~02:49} & B2-C1 & 9.6 & 27.0 & 0.46\\
\textcolor{orange}{2011-05-09~~03:14} & B2-C1 & 9.5 & 31.0 & 0.49\\
\textcolor{orange}{2011-05-09~~03:29} & B2-C1 & 9.4 & 33.4 & 0.52\\
\textcolor{green}{2011-05-09~~04:54} & B2-D0 & 25.9 & 46.4 & 0.18\\
\textcolor{green}{2011-05-09~~05:27} & B2-D0 & 24.7 & 51.6 & 0.16\\
\textcolor{orange}{2011-05-09~~08:28} & B2-C1 & 5.3 & 80.9 & 0.61\\
\textcolor{green}{2011-05-10~~02:23} & B2-D0 & 29.3 & 23.5 & 0.20\\
\textcolor{green}{2011-05-10~~03:46} & B2-D0 & 27.7 & 36.7 & 0.19\\
\textcolor{orange}{2012-02-01~~09:04} & A1-B2 & 10.7 & 103.5 & 0.51\\
\textcolor{purple}{2012-02-24~~05:20} & D0-H0 & 63.8 & 37.3 & 0.06\\
\textcolor{purple}{2012-02-24~~05:30} & D0-H0 & 63.8 & 39.8 & 0.06\\
\textcolor{purple}{2012-02-24~~09:29} & D0-H0 & 59.9 & 92.0 & 0.05\\
\textcolor{green}{2012-03-26~~09:05} & I1-K0 & 32.4 & 38.6 & 0.11\\
\textcolor{green}{2012-03-26~~09:41} & I1-K0 & 30.8 & 43.1 & 0.10\\
\textcolor{red}{2012-03-27~~05:21} & I1-K0 & 38.0 & 9.1 & 0.10\\
\textcolor{red}{2012-03-27~~05:26} & I1-K0 & 37.9 & 9.7 & 0.09\\
\textcolor{red}{2012-03-27~~05:34} & I1-K0 & 37.8 & 10.9 & 0.09\\
\textcolor{red}{2012-03-27~~07:09} & I1-K0 & 36.2 & 23.8 & 0.10\\
\textcolor{red}{2012-03-29~~04:08} & I1-K0 & 38.2 & 179.9 & 0.09\\

\hline
\end{tabular}
\end{center}

\tablefoot{For clarity, observations are color-coded by projected
baseline length, using the same color scheme as
Fig.~\ref{fig_uvplot}.}

\end{table*}

We used the prism as the dispersive element, resulting in
spectrally-resolved visibility amplitudes and differential phases with
a resolution of $\lambda / \Delta \lambda = 35$ in the wavelength
range $8-13$~\micron.  All observations were carried out in the
HIGHSENS mode, meaning the correlated flux and the photometric
measurements are done separately.  For a detailed description of this
observing procedure, see the work by \citet{Leinert04}.  In general,
every science measurement was preceded and/or followed by a
calibration measurement of a bright single star of known diameter.

The raw data were reduced using the MIA+EWS package version 1.7.1.
Masks were generated for each interferometric measurement from the
corresponding photometric data.  From the calibrator measurements
taken throughout the night, we computed the mean sensitivity function,
relating counts/s to Jy of correlated flux.  The uncertainty in this
sensitivity function was taken to be the standard deviation over all
the calibrator measurements in the night, and is typically on the
order of 10\%.

Because the quality of the photometric data from the ATs is generally
poor, we assume AFGL~4176 to be non-variable and use an average total
spectrum from the UTs for computing the visibility amplitude from the
correlated flux.  Uncertainties in the final visibilities were
calculated using standard error propagation formulae.

\subsection{Far-infrared imaging}

The AFGL~4176 region was covered by far-infrared observations taken by
the Herschel Space Observatory \citep{Pilbratt10}, within the
framework of the Hi-GAL survey \citep{Molinari10}.  In this paper, we
present bolometer data at 70 and 160~\micron, obtained with the PACS
instrument \citep{Poglitsch10} on August 16, 2010.  These data are
scan maps with a scan rate of 60$''$/s, which slightly degrades the
spatial resolution compared to the theoretical diffraction limit.
After onboard preprocessing, the data are transmitted with a downlink
rate of 5 frames/s for the 70~\micron{} filter, and 10 frames/s for
the 160~\micron{} filter.  The final resulting spatial resolution for
the 70~\micron{} data is about 9\farcs2, and 12\farcs6 for the
160~\micron{} data, although the true PSF is asymmetric.  For a more
detailed description of the observing procedure used in the Hi-GAL
program, we refer to the description of the observing strategy by
\citet{Molinari10}.

We extracted the level-0 data from the Herschel Science Archive of the
relevant $2\degr \times 2\degr$ strip covering AFGL~4176 (OBSIDs
1342203085 and 1342203086).  Despite the occurrence of very bright
far-infrared sources in the Hi-GAL fields, these observations were
performed with the normal high-gain settings for the PACS bolometers.
The level-0 data were processed with the HIPE software \citep{Ott10}
version 8.0.3178 up to the so-called level-1, where internal data
units have been translated to Janskys, and outliers (i.e. glitches) in
the data and saturated pixels have been removed or masked.

In addition to these standard procedures, we applied a task for
eliminating the electronic cross-talk within the bolometers, which can
cause the appearance of ghosts close to strong compact sources.
Furthermore, we use a module recently developed by the instrument team
which applies a non-linearity correction to the data.  This is
especially important for our purpose, since in the high-gain mode used
for the Hi-GAL program, very bright sources will trigger a non-linear
response in the bolometer behavior, resulting in an altered PSF and
decreasing measured fluxes by $10-15\%$.

The removal of $1/f$ noise and effects resulting from temperature
drifts within the bolometers, as well as the final mapping, was done
using the version~7 of the Scanamorphos program \citep[][submitted to
  PASP]{Roussel12}, for which we used the ``galactic,'' ``noglitch''
and ``parallel'' options.  Final maps were produced with a pixel scale
of 3\farcs2 for both filters, which is the native plate scale for the
70~\micron{} bolometers, and half the native plate scale for the
160~\micron{} bolometers.

\subsection{870~\micron{} imaging}

AFGL~4176 was also imaged at 870~\micron{} as part of the APEX
Telescope Large Area Survey of the Galaxy (ATLASGAL).  We refer to the
publication by \citet{Schuller09} regarding the general outline of
this sub-millimeter continuum survey, and the technical details of the
observations and the data reduction.  In the present work, we use a
map from the re-reduction of the data in 2011, done by the
F.~Schuller, using refined methods for thermal drift removal within
the bolometer reduction software BOA \citep{Schuller09}.  Furthermore,
the treatment of strong sources within the iterative process of the
data reduction has been improved.  Finally, the data have been mapped
onto a 6-arcsec grid, using version 2.18 of the mapping software SWarp
\citep{Bertin10}, which prevents the formation of spurious Moir\'e
patterns still present in earlier ATLASGAL maps.

\section{Results}
\label{sec_results}

\begin{figure}[ht!]
  \begin{center}
    \includegraphics[width=75mm]{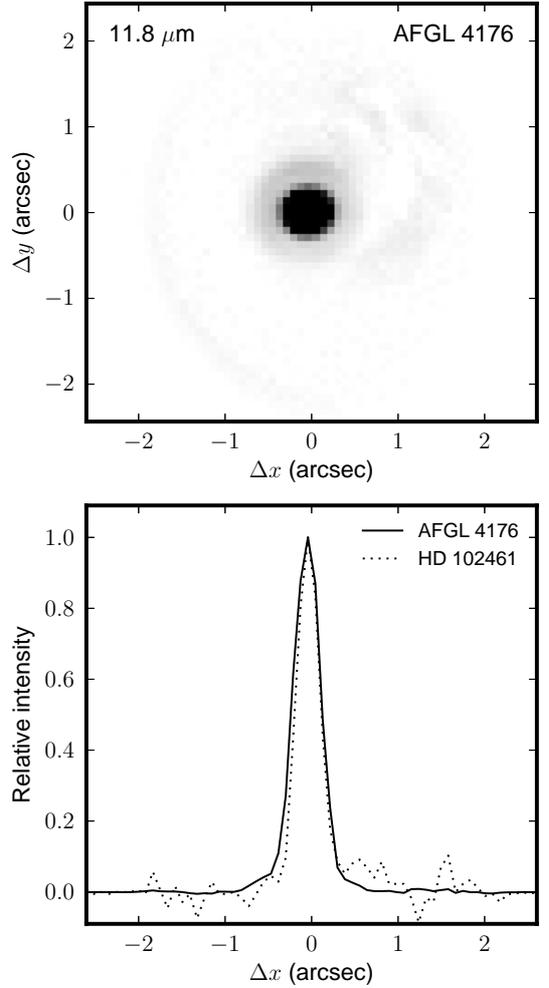}
    \caption{\textit{Above:} Acquisition image of AFGL~4176 from MIDI
      obtained on the 8.2~m UT3 telescope using the SiC filter, which
      has a central wavelength of 11.8~\micron{} and a width of
      2.32~\micron{}.  \textit{Below:} Comparison of a cut through the
      acquisition image of AFGL~4176, and an analogous image of the
      unresolved calibrator star HD~102461.}
    \label{fig_midiacq}
  \end{center}
\end{figure}

\begin{figure}[hb!]
  \begin{center}
    \includegraphics[width=80mm]{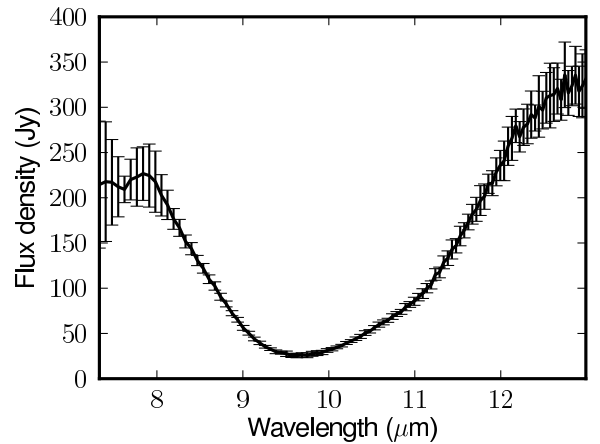}
    \caption{Average $N$-band spectrum measured with MIDI on the UTs.}
    \label{fig_midispect}
  \end{center}
\end{figure}

\begin{figure*}[ht!]
  \begin{center}
    \includegraphics[width=170mm,bb=90 108 520 684]{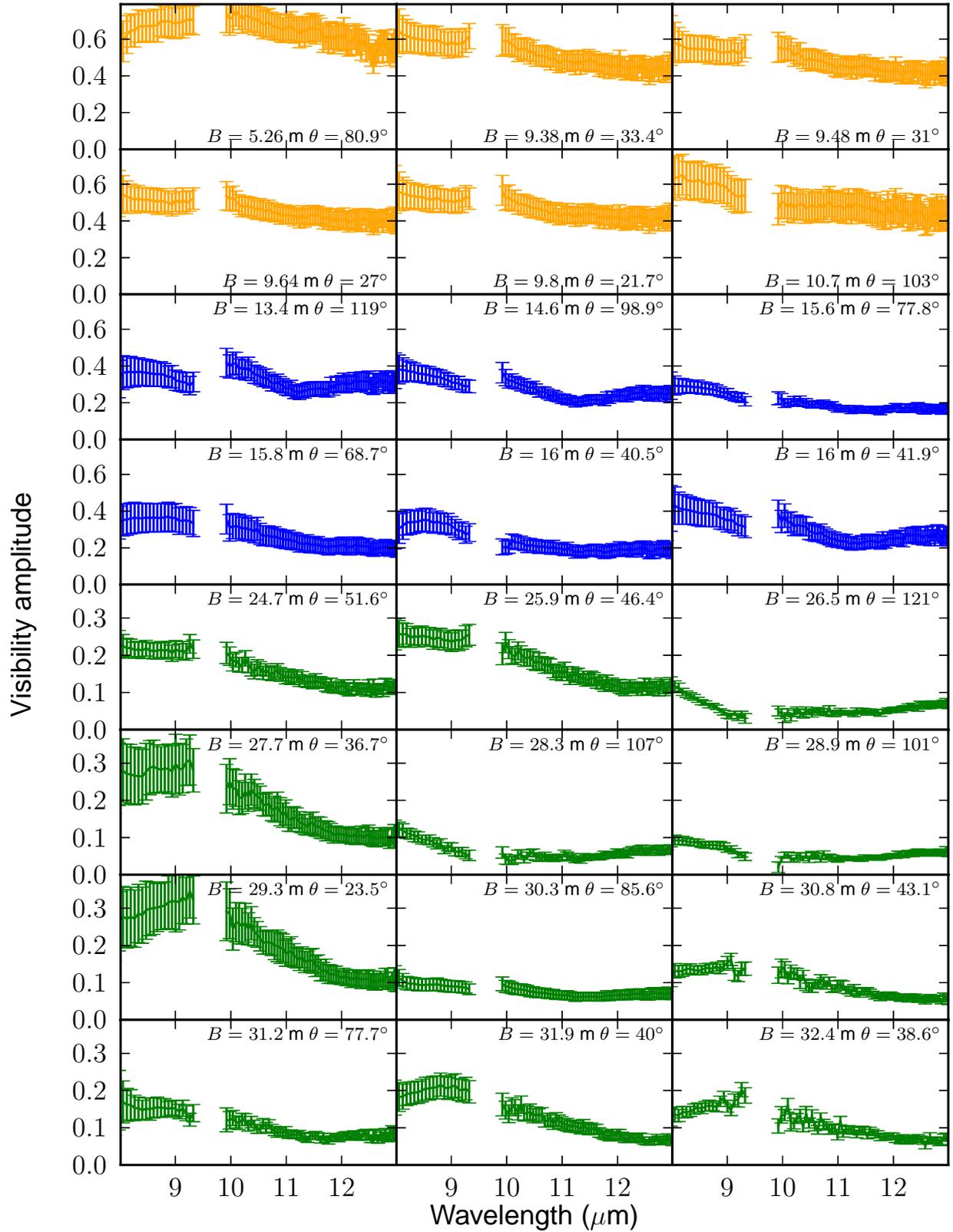}
    \caption{The complete set of visibility amplitudes measured with
      MIDI.  The plots are sorted by projected baseline length,
      indicated as $B$ in the inset text, together with the position
      angle $\theta$.  The color coding is the same as for
      Fig.~\ref{fig_uvplot}.}
    \label{fig_visamp}
    \addtocounter{figure}{-1}
  \end{center}
\end{figure*}

\begin{figure*}[ht!]
  \begin{center}
    \includegraphics[width=170mm,bb=90 298 520 684]{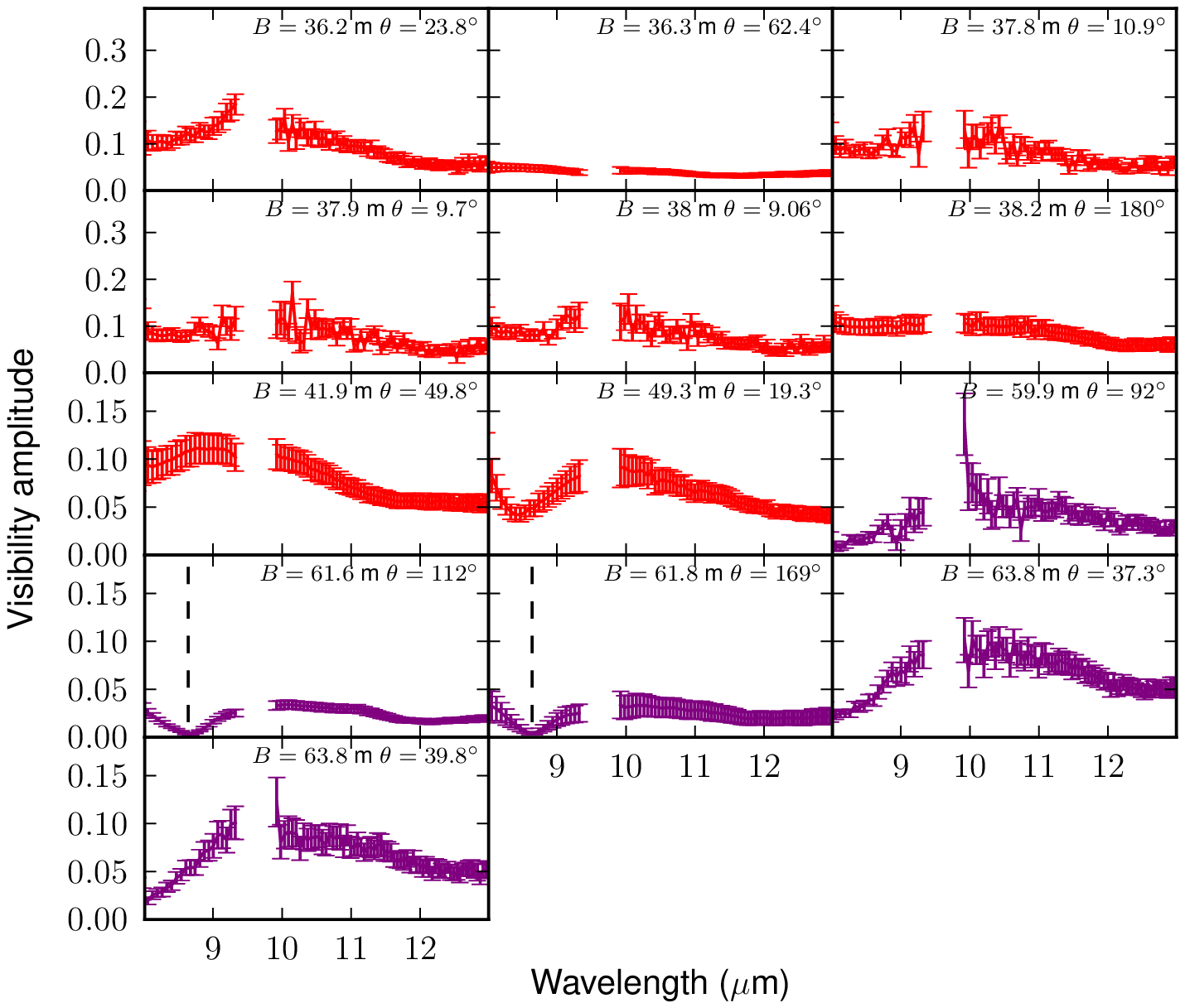}
    \caption{Continued.}
  \end{center}
\end{figure*}

\begin{figure*}[ht!]
  \begin{center}
    \includegraphics[width=170mm,bb=90 298 520 684]{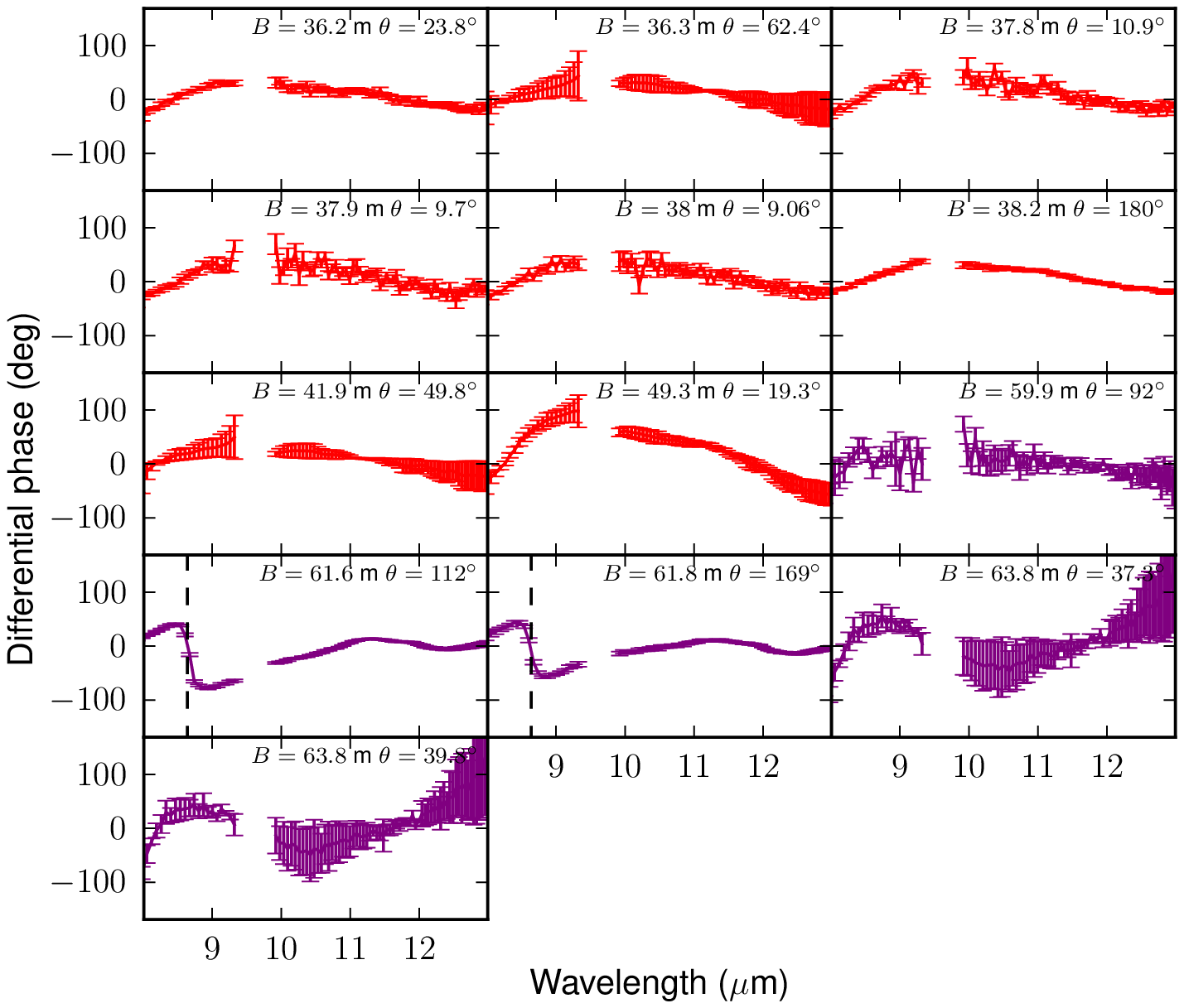}
    \caption{Differential phases measured with MIDI for $uv$ points
      with projected baselines longer than 36~m.  The vertical dashed
      line for the two measurements with $B=61.6$ and $61.8$~m shows
      the location of the phase flip, indicative of a zero crossing in
      the visibility amplitude.}
    \label{fig_visphi}
  \end{center}
\end{figure*}

We show an AO-corrected acquisition image of AFGL~4176 obtained with
the 8.2~m telescope UT3 on Mar.~2, 2005 in the upper panel of
Fig.~\ref{fig_midiacq}.  The image was obtained using the SiC filter,
which has a central wavelength of 11.8~\micron{} and a width of
2.32~\micron{}.  In the lower panel, we compare a cut through the
acquisition image of AFGL~4176 with an analogous cut through the
acquisition image of the unresolved calibrator star HD~102461, which
was observed immediately after the science source.  From the
similarity of the spatial cut through the image of AFGL~4176 to that
of the calibrator, we conclude that AFGL~4176 is at most marginally
resolved in diffraction-limited 8~m-class telescope images at $N$-band
wavelengths.

In Fig.~\ref{fig_midispect}, we show the average $N$-band spectrum of
AFGL~4176 from the 8.2~m UT telescopes, used to derive the visibility
amplitude from the correlated flux measurements.  The visibility
amplitudes as a function of wavelength are shown in
Fig.~\ref{fig_visamp}, where the measurements have been ordered by
projected baseline length, and are presented using the same color
scheme as in Fig.~\ref{fig_uvplot}.  Finally, in
Fig.~\ref{fig_visphi}, we show the differential phases measured with
MIDI for the measurements with $B > 36$~m (the differential phases at
shorter baselines are not shown, as they are essentially zero).

We note the clear presence of a zero crossing in both the visibility
amplitude and differential phase for the two measurements with
$B=61.6$ and $61.8$~m, indicated with a vertical dashed line in
Figs.~\ref{fig_visamp} and \ref{fig_visphi}.  We previously
interpreted this as evidence for circular symmetry at the smallest
scales probed \citep{Boley11}.  In order to test this hypothesis, we
targeted this spatial frequency for follow-up observations with the
ATs at other position angles on Feb.~24, 2012 (the remaining
``purple'' measurements in Figs.~\ref{fig_uvplot}, \ref{fig_visamp}
and \ref{fig_visphi}).  However, while a non-zero differential phase
is measured, a zero crossing is clearly \emph{not} present in these
follow-up observations, implying deviations from spherical symmetry at
the smallest angular scales probed by MIDI.

\begin{figure*}[ht!]
  \begin{center}
    \includegraphics[width=170mm]{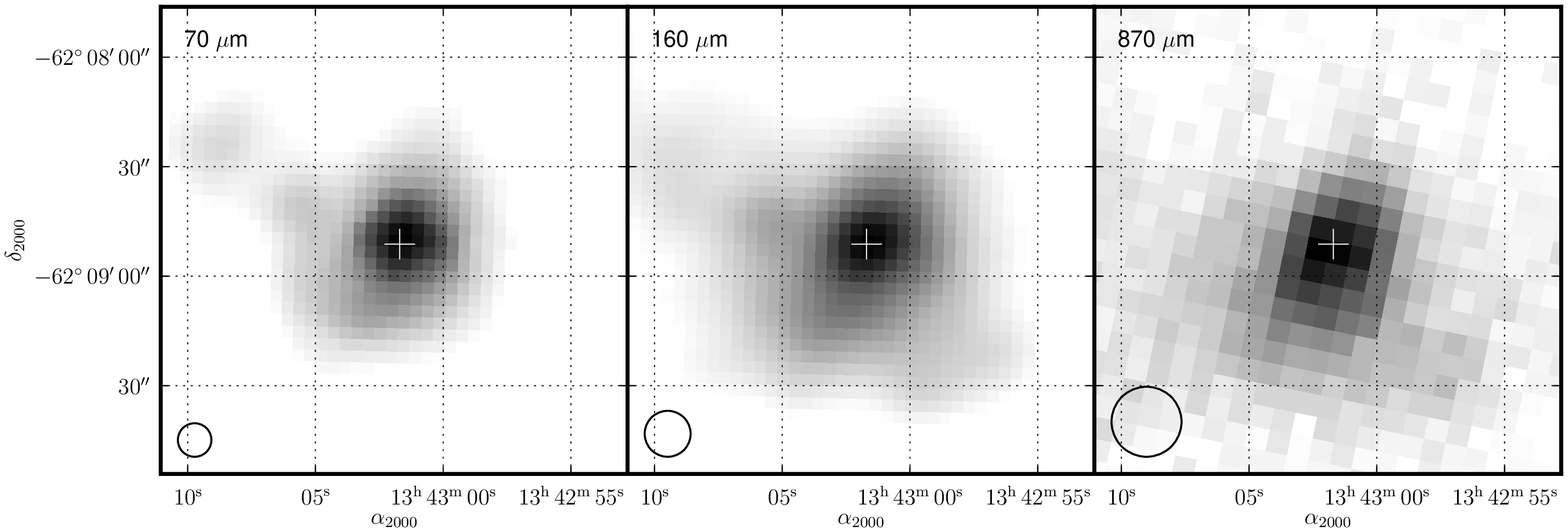}
    \caption{The 70 and 160~\micron{} images obtained from the Hi-GAL
      survey with the Herschel Space Observatory, and the
      870~\micron{} image from the ATLASGAL survey with the APEX
      telescope.  The circle in the lower left corner of each image
      indicates the beam size.  The two Hi-GAL images are shown with
      logarithmic scaling of the intensity, while the ATLASGAL image
      is shown with linear scaling.}
    \label{fig_images}
  \end{center}
\end{figure*}

In Fig.~\ref{fig_images}, we show the images at 70 and 160~\micron{}
(Hi-GAL) and 870~\micron{} (ATLASGAL).  The approximate beam size for
each image is 9\farcs2, 12\farcs6 and 19\farcs2, respectively, and is
shown as a circle in the lower left-hand corner of each image.  The
position of the infrared source is indicated as a cross in each image.
The position of the emission peak in the Hi-GAL data corresponds to
that of the 2MASS source to within the pointing accuracy of the
observations.  In the ATLASGAL data, the emission peak lies
$\sim4$\arcsec{} south of the IR position, although this is also
within the rms pointing accuracy of the ATLASGAL observations.

At the Hi-GAL wavelengths, within a 36\arcsec{} aperture, we measure
flux densities of 2382~Jy and 1383~Jy at 70~\micron{} and
160~\micron{}, respectively.  While the absolute flux calibration on
these data is accurate to within 5\% (Herschel PACS Report
PICC-ME-TN-037, Apr. 12, 2011), there is significant, non-uniform
background emission in the region which is difficult to account for.
Therefore, these values should be treated as upper limits, as it is
not clear where the background begins to dominate and the extended
emission associated with the source stops.  The 870~\micron{} emission
is also quite extended; within a 19\arcsec{} aperture, which covers
just the central emission peak, we measure an integrated flux density
of 2.86~Jy, while in a larger 72\arcsec{} aperture, covering the
extended emission, we measure an integrated flux density of 10.6~Jy.

\begin{table}[htpb]
  \begin{center}
    \caption{Photometric data}
    \label{tab_photdata}
    \begin{tabular}{l l c c}
      \hline \hline
      Wavelength & Flux density & Aperture & Reference \\
      (\micron) & (Jy) & (\arcsec) & \\
      \hline
      1.25 & $(2.664 \pm 0.139) \times 10^{-3}$ & 4 & (1) \\
      1.65 & $0.123 \pm 0.004$ & 4 & (1) \\
      2.20 & $1.760 \pm 0.04$ & 4 & (1) \\
      9.0 & $173.2 \pm 1.33$ & 11 & (2) \\
      12.0 & $250.70 \pm 12.535$ & $-$ & (3) \\
      18.0 & $434.2 \pm 13.3$ & 11 & (2) \\
      25.0 & $601.60 \pm 30.080$ & $-$ & (3) \\
      60.0 & $2701.00 \pm 243.09$ & $-$ & (3) \\
      65.0 & $1506.0 \pm 434.0$ & 27 & (2) \\
      70.0 & $2382.2$ & 36 & (4) \\
      90.0 & $849.1 \pm 208.0$ & 27 & (2) \\
      100.0 & $3566.00 \pm 534.90$ & $-$ & (3) \\
      140.0 & $1437.0 \pm 127.0$ & 44 & (2) \\
      160.0 & $1939.0 \pm 293.0$ & 44 & (2) \\
      160.0 & $1383.4$ & 36 & (4) \\
      870.0 & $2.86$ & 19 & (4) \\
      870.0 & $10.6$ & 72 & (4) \\
      1200.0 & $3.86$ & 39 & (5) \\
      \hline
    \end{tabular}
  \end{center}
  \tablebib{(1)~2MASS; (2)~AKARI; (3)~IRAS; (4)~This work;
    (5)~\citet{Beltran06}}
\end{table}

\begin{figure*}[htpb]
  \begin{center}
    \includegraphics[width=170mm]{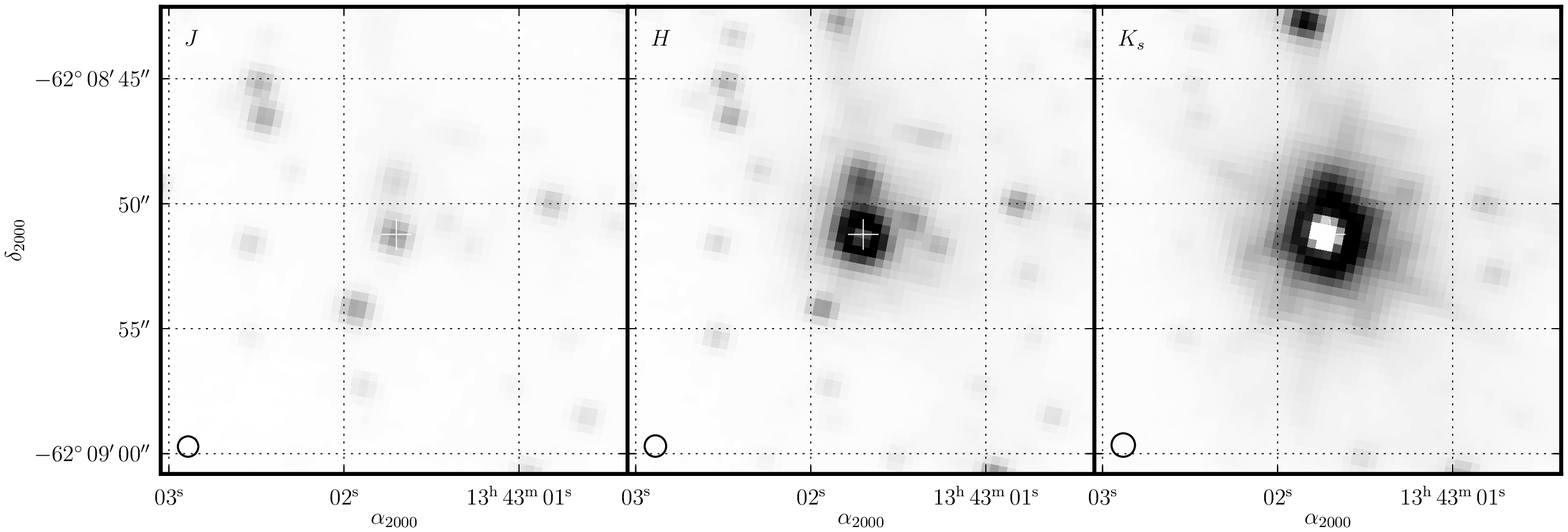}
    \caption{$J$-, $H$- and $K_s$-band images from the VVV survey.
      AFGL~4176, indicated with a white cross, is saturated in both
      the $H$ and $K_s$ images.  The circle in the lower left corner
      of each image indicates the seeing, as measured from the guide
      star by the autoguider.}
    \label{fig_images_VVV}
  \end{center}
\end{figure*}

\begin{figure}[!h]
  \begin{center}
    \includegraphics[width=85mm]{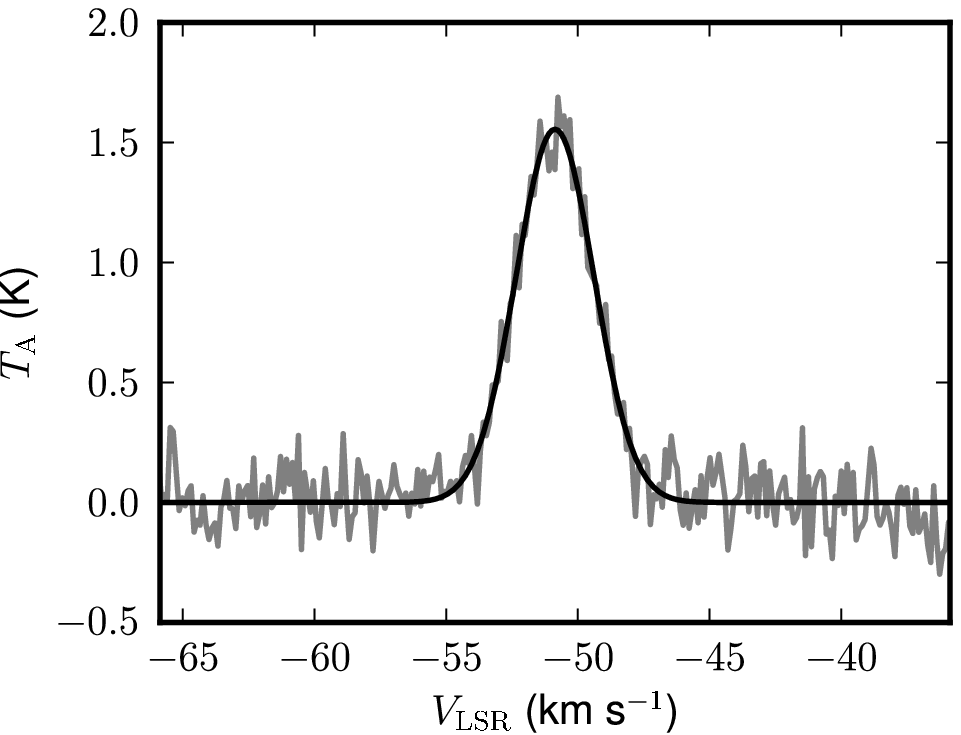}
    \caption{Spectrum of HCO+ from the MALT90 survey.  The
      observations are shown with the gray line, while a Gaussian fit
      is shown with the black line.}
    \label{fig_hcopspect}
  \end{center}
\end{figure}

In Table~\ref{tab_photdata} we list the photometric data and effective
aperture sizes from previous works and surveys which we consider in
modeling our source.  Unless noted otherwise, we compare our model
data (see Sec.~\ref{sec_radtran_modeling}) to the observations by
considering the flux contribution within a Gaussian aperture with the
same FWHM as the observational data.

In Fig.~\ref{fig_images_VVV}, we show $J$-, $H$- and $K_s$-band images
of AFGL~4176 and its immediate surroundings from the first data
release of the ESO public survey VISTA Variables in the V\'{i}a
L\'{a}ctea \citep[VVV;][]{Saito12}.  These images were obtained on the
4~m VISTA telescope on Cerro Paranal, and surpass 2MASS both in
spatial resolution and sensitivity.  The seeing for the $J$, $H$ and
$K_s$ images presented here is 0\farcs82, 0\farcs86 and 0\farcs94,
respectively.  AFGL~4176 is present as a bright, compact source, and
saturated in both the $H$ and $K_s$ images.

Finally, the region around AFGL~4176 was mapped in several molecular
lines as part of the Millimeter Astronomy Legacy Team 90~GHz (MALT90)
survey \citep{Foster11} with the Mopra 22~m telescope.  Of the 16
spectral lines which were mapped as part of the survey, HCO+ was the
brightest in AFGL~4176.  The HCO+ line data, shown in
Fig.~\ref{fig_hcopspect}, have a spectral resolution of $\sim
0.11$~km~s$^{-1}$.  This line is a tracer of dense molecular gas, and
has been used as an outflow indicator \citep[e.g.][]{Hofner01}.
However, in the case of AFGL~4176, the spectrum is very well
approximated by a Gaussian (FWHM 3.5~km~s$^{-1}$), and shows no signs
indicating the presence of an outflow.

\section{Analysis}

\subsection{One-dimensional radiative transfer modeling}
\label{sec_radtran_modeling}

As shown in Sec.~\ref{sec_results}, AFGL~4176 shows evidence for
deviations from spherical symmetry only at the scales probed by our
mid-infrared interferometric observations, which correspond to scales
of roughly 20 to 200 mas.  Furthermore, no outflow component has been
revealed in searches by \citet{deBuizer03} and \citet{deBuizer09}, nor
in the HCO+ data presented in this work (Fig.~\ref{fig_hcopspect}).
Since there is no evidence for asymmetry or preferred geometry on
scales larger than a few hundred AU, we begin with one-dimensional
models to explore the overall structure of this source.

We model AFGL~4176 using the one-dimensional radiative transfer code
MoDust, which was developed by A. de~Koter and J. Bouwman.  MoDust
uses the Feautrier method to solve the radiative transfer equation
\citep{Feautrier64}, and has been used previously by
e.g. \citet{Malfait99,Kemper01,Bouwman01}.  Using MoDust, the dust
temperature structure is iteratively solved for a given density
structure and dust properties.  Different opacities can be given for
different temperature regimes, thus making it possible to dynamically
determine the distributions of different types of dust throughout the
model.

Using this code, we produce radial intensity profiles of the source
image on the sky at all wavelengths for which we have observations,
i.e. from 1.25~\micron{} to 1.2~mm.  From these radial intensity
profiles, we calculate aperture-corrected flux values for the SED,
PSF-convolved images for comparison with the Hi-GAL and ATLASGAL
data\footnote{For the PACS array on Herschel, we took the PSFs from
  the PACS calibration web site
  (http://herschel.esac.esa.int/twiki/bin/view/Public/PacsCalibrationWeb).
  For the LABOCA bolometer array used on the APEX telescope, we take
  the PSF to be well-approximated by a Gaussian with FWHM = 19\farcs2
  \citep{Siringo09}.}, and interferometric visibilities for comparison
with the MIDI data.  We do not attempt to match the near-IR VVV
images, as AFGL~4176 is saturated in them.  The relative goodness of
fit of models was assessed by constructing a $\chi^2$ estimate which
includes roughly equal contribution from these three measures.

\subsubsection{Constraints and uncertainties on fundamental parameters}

As noted in the introduction, the distance to AFGL~4176 cannot be
reliably determined at present.  We therefore chose to investigate
both near and far models, at distances of 3.5 and 5.3~kpc,
respectively.

Under the assumption that the \ion{H}{II} region seen in radio
observations is powered by the AFGL~4176 source, the number of
ionizing photons, and consequently a minimum inferred spectral type,
can be determined.  The radio flux measurements at 843 and 1415~MHz of
\citet{Caswell92} appear to be consistent with thermal bremsstrahlung
radiation from an \ion{H}{II} region.  For a distance of 5.3~kpc,
using the main-sequence luminosities for O stars from
\citet{Martins05} and the stellar atmosphere models of
\citet{Kurucz79}, we find a minimum spectral type of the central
source of about O8.  This corresponds to an effective temperature of
about 33\,000~K, which we adopt for the temperature of the central
blackbody source in our models.

For the amount of foreground extinction, we adopt a simple
prescription of 1 magnitude of visual extinction per kiloparsec,
giving us $A_V = 3.5$ and $5.3$~mag for the near and far models,
respectively.  This small amount of additional (i.e. interstellar)
extinction was applied on top of our models, and has the effect of
further lowering the near-IR flux levels and slightly deepening the
10~\micron{} silicate feature.

\subsubsection{Dust model}

One of the foremost practical constraints on the choice of dust model
is the need for opacities at short wavelengths ($\lambda \leq
0.1$~\micron).  Since the luminosity of this source is on the order of
$10^5$~\Lsun{} and the amount of ionizing flux (inferred from the
radio continuum observations) is equivalent to at least an O8 star
star, the vast majority of the stellar radiation is emitted at UV
and optical wavelengths.

A number of attractive dust models for conditions expected in
protostellar cores, envelopes and protoplanetary disks have been
developed \citep[for example by][]{Ossenkopf94,Henning96,Semenov03}.
However, a common shortcoming of these models is the lack of opacity
information at short (UV) wavelengths, due to convergence problems of
the grain models (D. Semenov, private communication).  On the other
hand, the empirically derived opacities of \citet{Laor93} and
subsequent works cover wavelengths from 100~\AA{} through 1~mm, with
the caveats that the real physical makeup of the dust is unknown, and
that at far-IR and (sub)mm wavelengths, where extinction is
effectively impossible to measure, the dust opacities are not well
known.

We therefore adopt a multi-component, temperature-dependent dust
model, guided by the presence of known spectral features in the
observations, and limited by the available coverage and reliability of
different opacity models at different wavelength regimes.  The
transition between different regimes, and therefore the presence of
different species throughout the surrounding material, are governed by
the temperature structure iteratively solved by MoDust.

Based on the presence of the 3.1~\micron{} water ice absorption
feature in the ISO SWS spectrum, we infer that at least some of the
amorphous silicate grains are surrounded by ice mantles, formed in the
cold material of the parent molecular cloud.  To include this in our
model, we use opacities from the coagulation model of
\citet{Ossenkopf94}.  Following the suggestions in that paper, we
interpolate between columns 1 and 2 of their Table~1 using the depth
of the 3.1~\micron{} ice feature.  From the ISO SWS spectra, we
measure $\tau_{3.1}=0.4$, and use the resulting interpolated opacities
for regions where $\Tdust < 130$~K.

For grains with $\Tdust > 130$~K, we use the ``astronomical
silicates'' and graphite grains of \citet{Laor93} with the ``MRN''
distribution of grain sizes ($n(a) \propto a^{-3.5}$ for grains
between 5~nm and 250~nm, where $n$ is the number of grains, and $a$ is
the grain radius), which is generally appropriate for diffuse ISM
conditions.  We adopt a ratio of 60:40 for the silicate to graphite
mass abundance in this temperature regime, although we note this is
not directly constrained by observations.  For the graphite grains, we
adopt a destruction temperature of 950~K, which corresponds to the
process of OH sputtering \citep{Duschl96}.  For the silicate grains,
we do not enforce a destruction temperature; at the inner edge of our
envelope models, the silicate grains typically reach a temperature of
about 2000~K.

The dust model therefore consists of a region of ``cold'' ($\Tdust <
130$~K) amorphous silicate and carbon grains with ice mantles, as well
as ``warm'' (130~K$< \Tdust < 950$~K for graphites, and 130~K$<
\Tdust$ for silicates) amorphous grains similar to the ``standard''
diffuse ISM mix.

\subsubsection{Density structure}

For the density structure of the dust in our model, we adopt a
piecewise power law consisting of concentric, spherical shells, with
no gaps in between the shells.  In order to simultaneously reproduce
the observed visibility levels, far-IR/sub-mm spatial distributions
and SED with this model, we find it necessary to include at least
three shells, with a jump in density at the interface between the
first (innermost) shell and the second shell.  At the interface
between the second and third shells, however, we were able to achieve
acceptable fits without a density jump, thereby avoiding the need for
an additional model parameter describing this.  We therefore
characterize the density of material in the outer shells by the
density $\rho_{23}$ at the contact point $r_{23}$ of the second and
third shells.  The precise structure used, therefore, has in principle
nine free parameters ($r_1$, $r_2$, $r_{23}$, $r_o$, $\rho_1$,
$\rho_{23}$, $p_1$, $p_2$, $p_3$) and follows the form
\begin{equation}
  \label{eq_density}
  \rho(r) = \left\{ \begin{array}{ll}
    \rho_{1} \left(\frac{r_1}{r}\right)^{p_1} & r_1 < r \leq r_2\\
    \rho_{23} \left(\frac{r_{23}}{r}\right)^{p_2} & r_2 < r \leq r_{23} \\
    \rho_{23} \left(\frac{r_{23}}{r}\right)^{p_3} & r_{23} < r \leq r_o
  \end{array} \right.
\end{equation}

\subsubsection{Results}

\begin{table*}
  \begin{center}
    \caption{One-dimensional radiative transfer model parameters}
    \label{tab_modelfit}
    \begin{tabular}{l c c c c c c c c c c c c}
      \hline \hline
      Model & $D$ & $r_1$ & $r_2$ & $r_{23}$ & $r_o$ & $\rho_1$ & $\rho_{23}$ & $p_1$ & $p_2$ & $p_3$ & $M_{\mathrm{dust}}$ & $L_\star$ \\
      & (kpc) & (AU) & (AU) & (AU) & (AU) & (g cm$^{-3}$) & (g cm$^{-3}$) & & & & (\Msun) & ($10^5$~\Lsun) \\
      \hline
      Near & 3.5 & 40 & 65 & 70\,000 & 125\,000 & $1.71 \times 10^{-18}$ & $3.56 \times 10^{-22}$ & 1.8 & 1.05 & 0.5 & 4.70 & 1.44 \\
      Far & 5.3 & 58 & 108 & 56\,000 & 195\,000 & $9.22 \times 10^{-19}$ & $4.71 \times 10^{-22}$ & 1.2 & 0.9 & 0.5 & 15.8 & 2.88 \\
      G91 & 4.0 & --- & 100 & 2646 & 50\,134 & --- & $1.49 \times 10^{-21}$ & --- & 2.0 & 0.0 & 1.32 & 1.57 \\
      \hline
      \end{tabular}
  \end{center}
\end{table*}

\begin{figure}[ht!]
  \begin{center}
    \includegraphics[width=85mm]{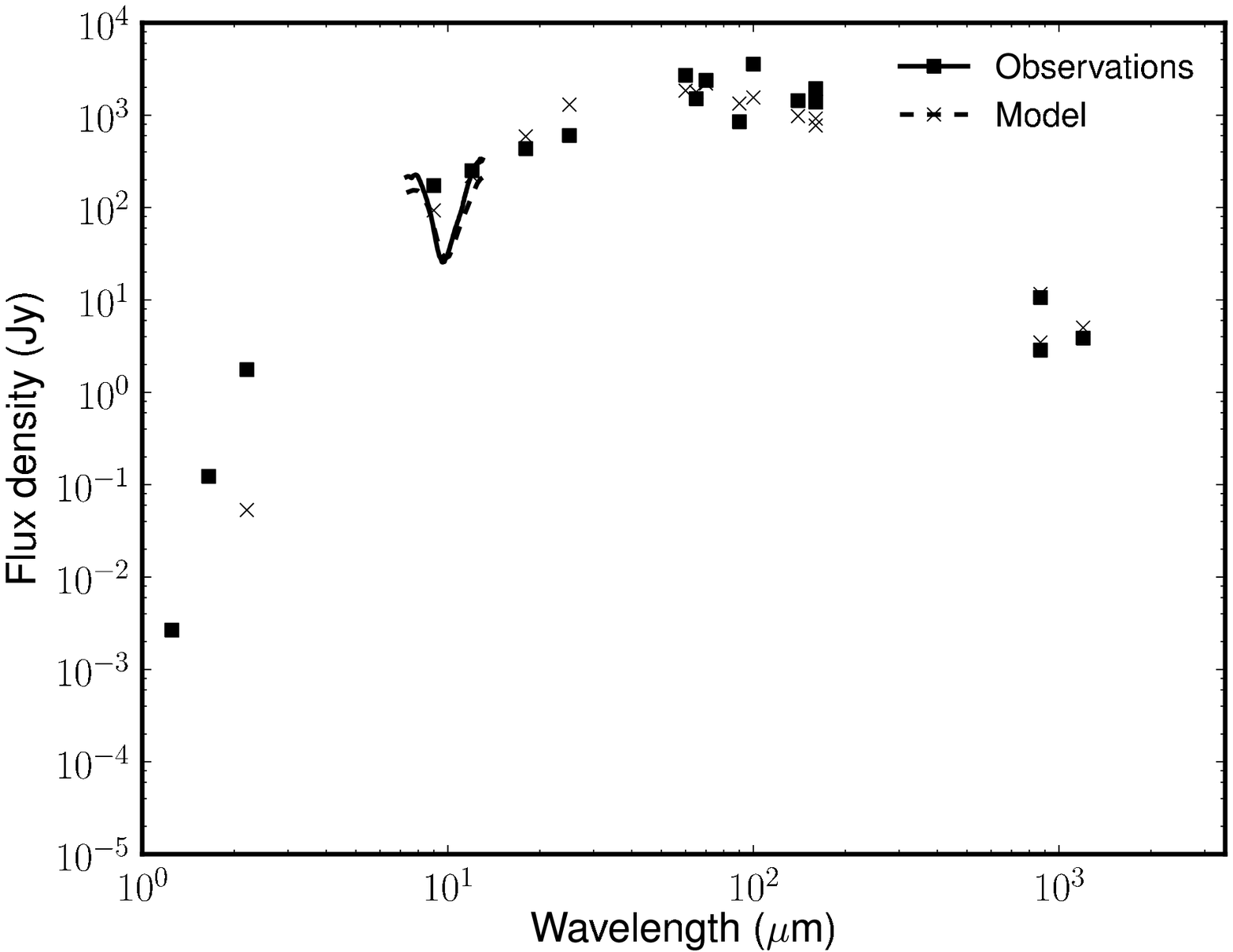}
    \includegraphics[width=85mm]{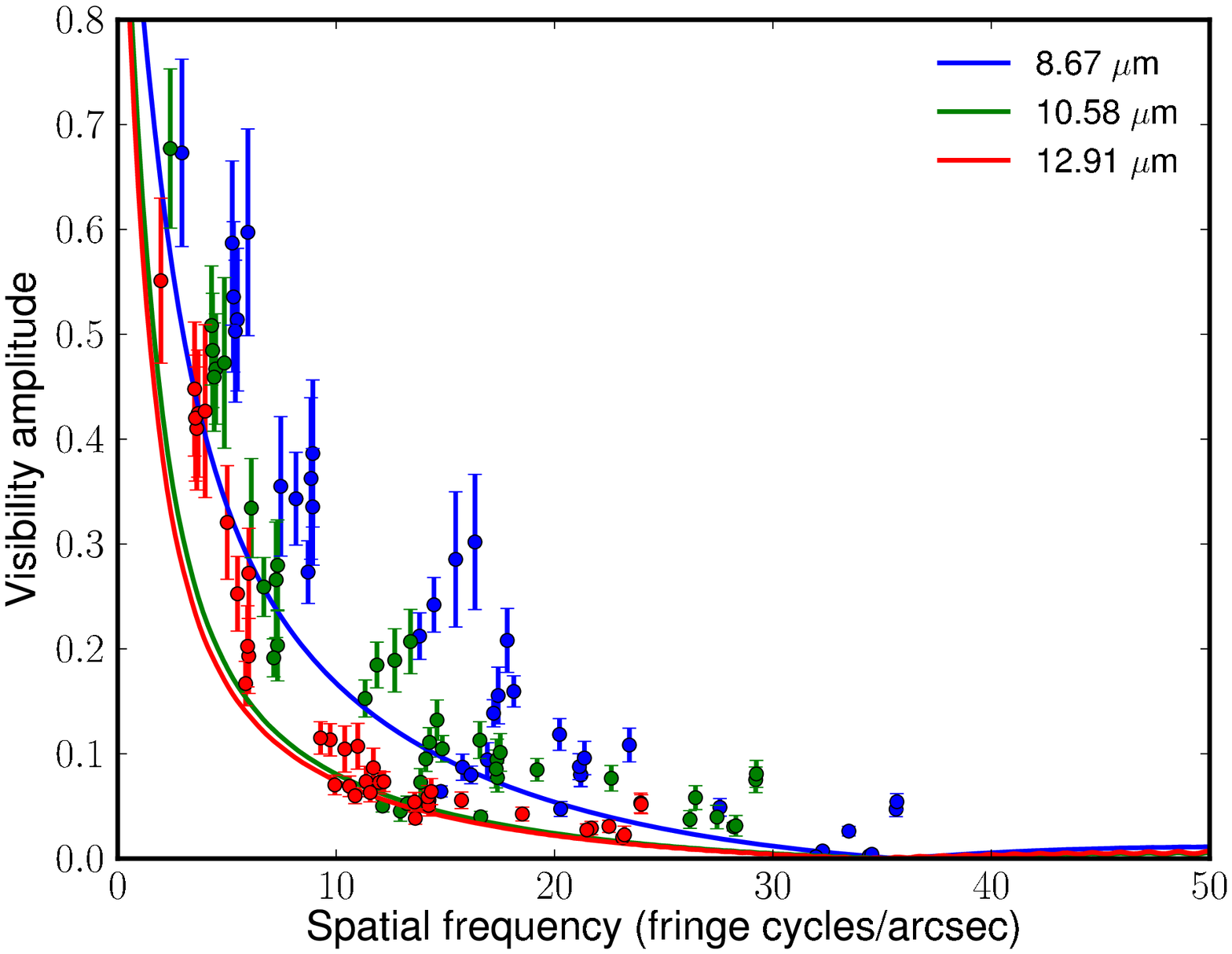}
    \includegraphics[width=85mm]{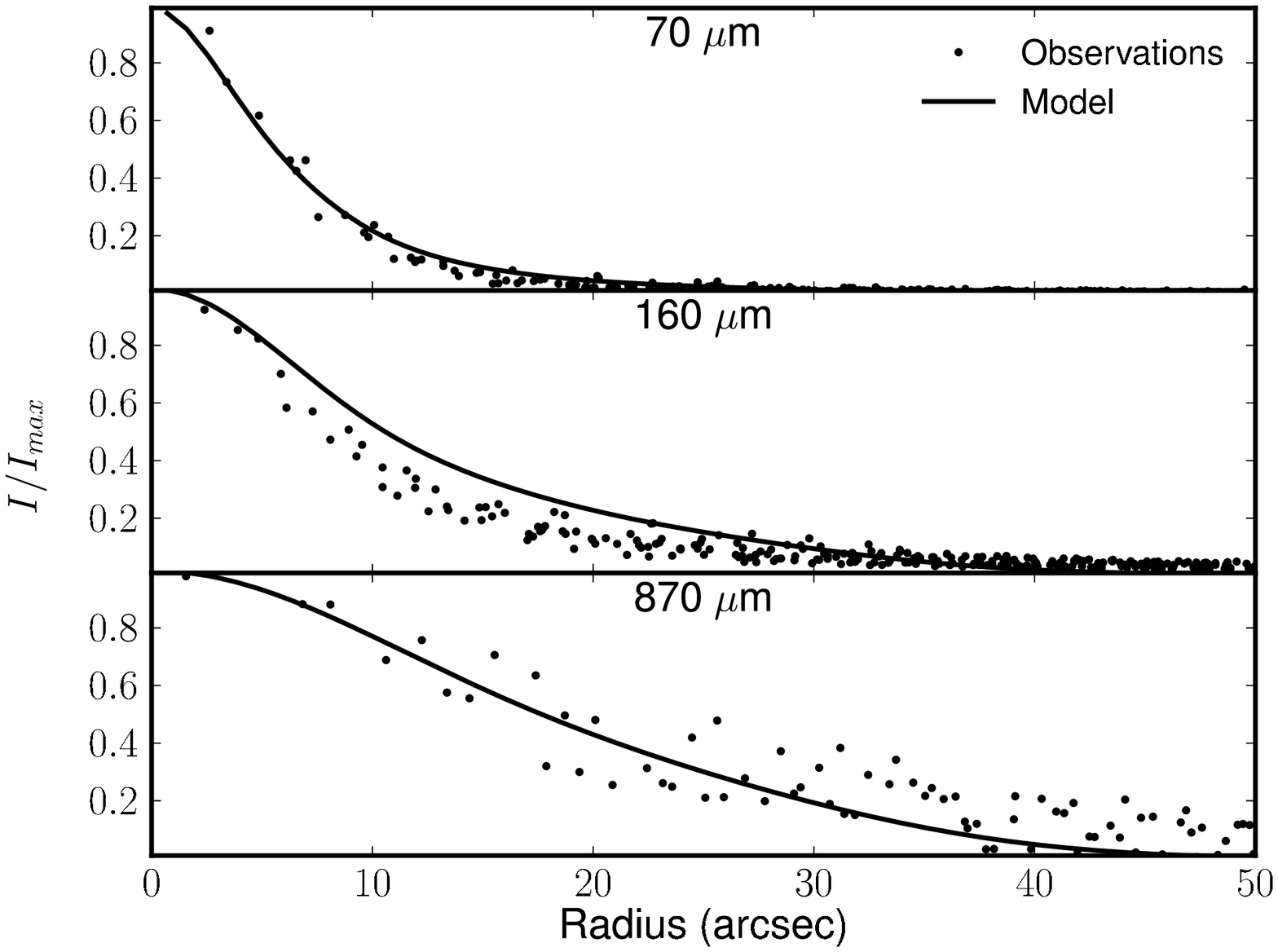}
    \caption{Model fits of the observational data for the near
      model. {\it Top:} The SED, where the aperture-corrected model
      flux levels are shown as crosses and the dotted line, and the
      observed flux values are shown as filled squares and the solid
      line. {\it Middle:} The modeled (lines) and observed (points)
      MIDI visibilities for three wavelengths. {\it Bottom:} The
      modeled (blue points) and observed (green points) radial
      intensity distributions from the Hi-GAL (70 and 160~\micron) and
      ATLASGAL (870~\micron) images.}
    \label{fig_nearfit}
  \end{center}
\end{figure}

\begin{figure}[ht!]
  \begin{center}
    \includegraphics[width=85mm]{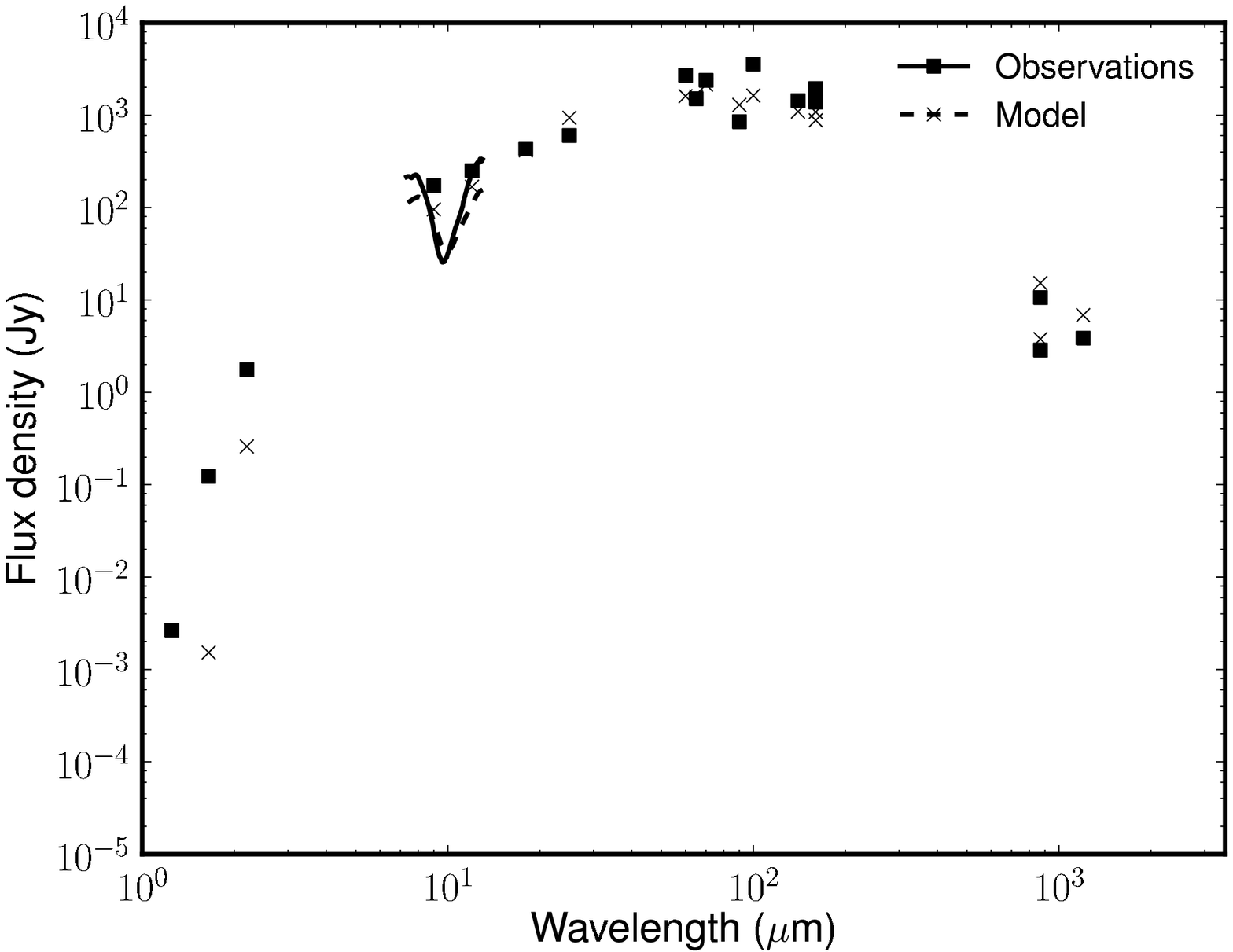}
    \includegraphics[width=85mm]{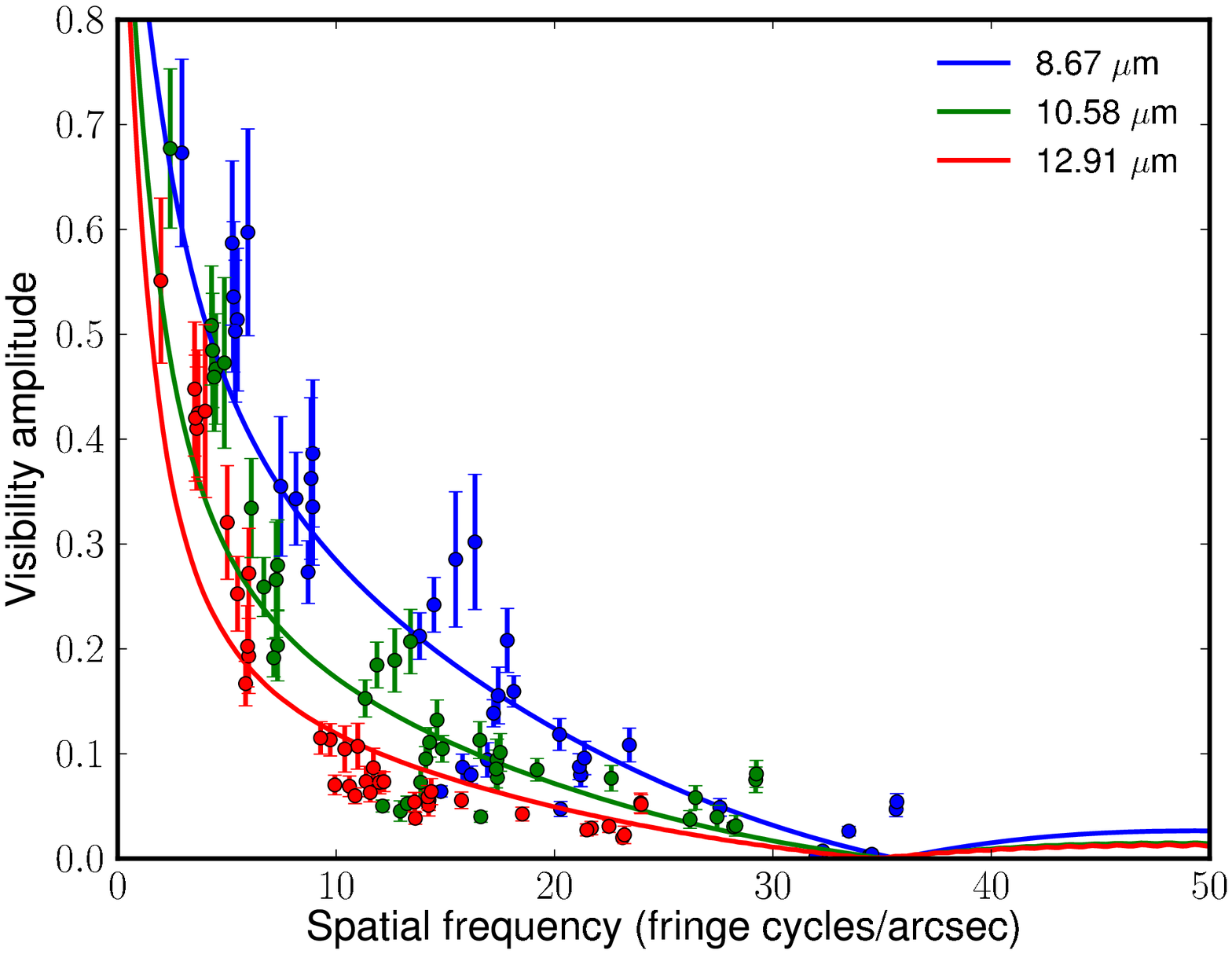}
    \includegraphics[width=85mm]{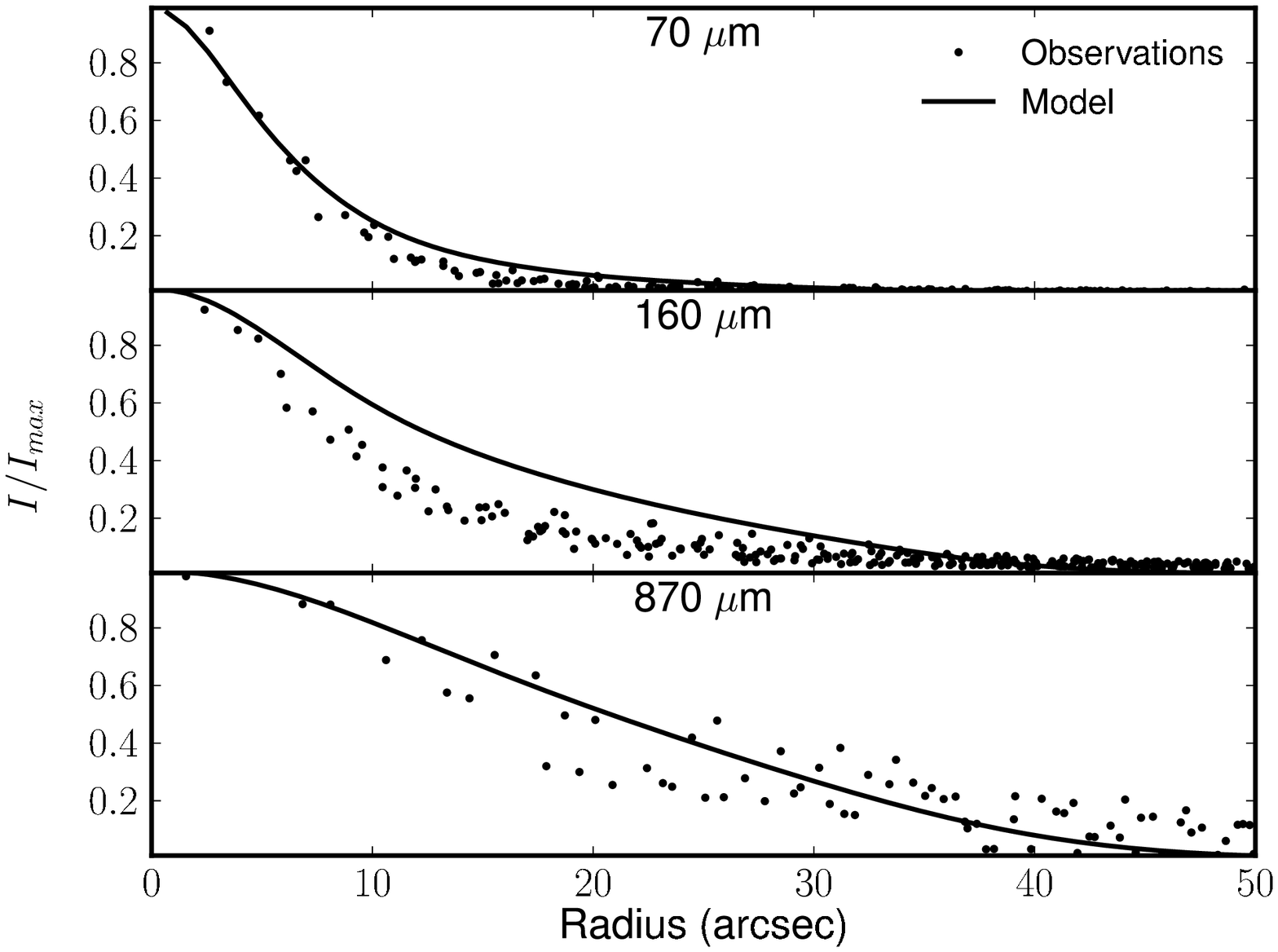}
    \caption{Model fits of the observational data for the far model.
      Figure notation is the same as for Fig.~\ref{fig_nearfit}.}
    \label{fig_farfit}
  \end{center}
\end{figure}

We present two fits to the observational data using the three-shell
model: one for the ``near'' distance of 3.5~kpc, and one for the
``far'' distance of 5.3~kpc.  The parameters are summarized in the
first two lines of Table~\ref{tab_modelfit}; comparisons with the
observations are shown in Figs.~\ref{fig_nearfit} and
\ref{fig_farfit}: the top panel shows the modeled flux levels from
the photometric data from Table~\ref{tab_photdata}; the center panel
shows the modeled MIDI visibilities at three wavelengths; and the
lower panel shows the radial distributions from Herschel (70 and
160~\micron{}) and APEX (870~\micron{}).

We defer a detailed discussion of the results to
Sec.~\ref{sec_discussion}, but note the most prominent points of the
model fits here.  Both models reproduce the SED reasonably well at
wavelengths longer than about 8~\micron{}, but significantly
underestimate the near-infrared flux levels.  Both models also
reproduce the \emph{overall} visibility levels, as well as the spatial
frequency of the zero crossing, however the far model provides a much
better match, particularly at the two shorter wavelengths (red and
green points in the middle panels of Figures~\ref{fig_nearfit} and
\ref{fig_farfit}).  Finally, despite the good agreement of the radial
profiles at wavelengths of 70 and 870~\micron{}, both models
overestimate the extent of emission at 160~\micron{}, although the
near model is in better agreement with the observed radial profile.
The 160~\micron{} filter might be contaminated by strong [\ion{C}{II}]
emission at 157~\micron{}, detected in emission in the ISO LWS
spectrum, which could alter the spatial profile of the brightness
distribution seen in this broadband filter.  However, since no
spatially-resolved [\ion{C}{II}] spectral map is available for
AFGL~4176, this ad-hoc explanation is not beyond doubt.

\subsection{Two-dimensional geometric modeling}
\label{sec_2dmodel}

\begin{figure}[h!]
  \begin{center}
    \includegraphics[width=85mm]{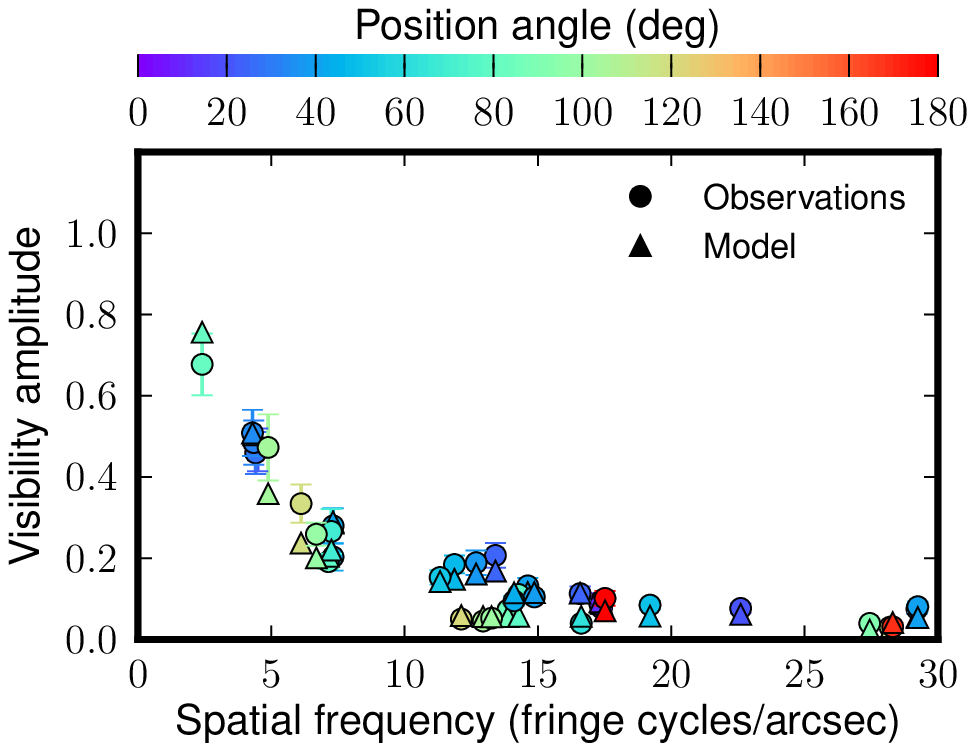}
    \caption{Geometric fit of a disk to the MIDI visibilities at
      10.6~\micron{}.  The color coding indicates the position angle
      of the measurements; circles show the observed values of the
      visibility amplitude, while triangles show the model values at
      the corresponding coordinates in $uv$ space.}
    \label{fig_2d_visfit}
  \end{center}
\end{figure}

\begin{figure}[h!]
  \begin{center}
    \includegraphics[width=85mm]{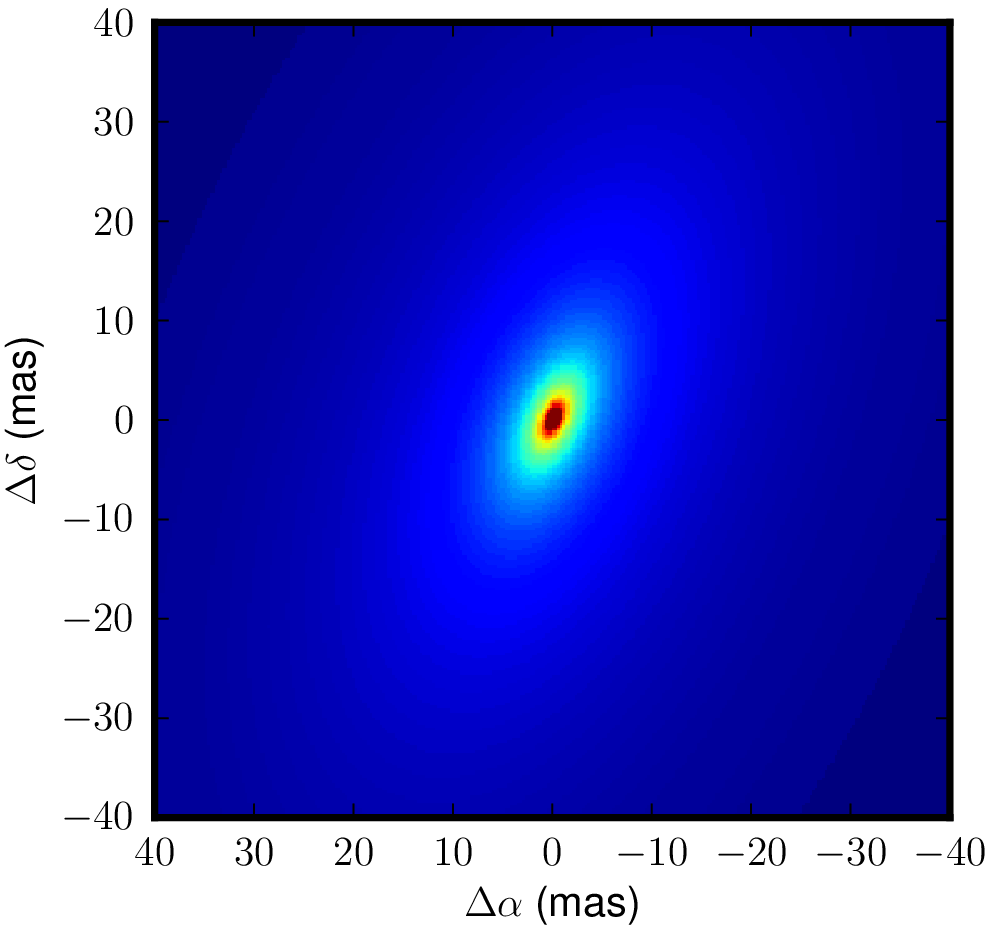}
    \caption{Image of the fit of the disk model to the mid-infrared
      visibilities at $\lambda=10.6$~\micron{}.}
    \label{fig_2d_image}
  \end{center}
\end{figure}

While the one-dimensional approaches used above can provide useful
insight, the deviations from spherical symmetry observed in the MIDI
data should be explored.  Such deviations can become especially clear
in cases where multiple $uv$ points have been measured with similar
projected baselines (i.e., spatial frequencies), but differing
position angles.  Any significant variations in the visibility
amplitude for a given spatial frequency will thus require deviations
from a circularly-symmetric intensity distribution to be explained.

We see clear signs of such behavior for projected baselines longer
than about 25~m, particularly for spatial frequencies larger than
$\sim 10$ fringe cycles/arcsec.  For small variations in spatial
frequency, we see visibility levels varying by over a factor of 5.
Furthermore, there is a clear dependence on position angle: for
measurements with $V < 0.2$, the visibility amplitude decreases with
increasing position angle, while for measurements with $V > 0.2$, the
visibility amplitude \emph{decreases} with increasing position angle.

This behavior (visibility amplitude increasing with position angle for
$V < 0.2$, and decreasing with position angle for $V > 0.2$) cannot be
modeled with a simple elongated Gaussian.  However, such a trend is
possible in the case of a source distribution which has a well defined
``edge'' in its intensity distribution, causing a zero-crossing in the
complex visibility with substantial power in the first side-lobe, for
example a disk model.

\begin{table}
  \begin{center}
    \caption{Two-dimensional geometric fit parameters}
    \label{tab_diskfit}
    \begin{tabular}{l c}
      \hline \hline
      Parameter & Best fit value \\
      \hline
      $r_o$  & 158~mas$^1$ \\
      $p$  & 0.467 \\
      $\phi$ & $158\degr$ \\
      $\theta$ & $59.7\degr$ \\
      $F_G/F_D$ & 1.64 \\
      $\theta_G$ & 143~mas$^2$ \\
      \hline
    \end{tabular}
  \end{center}
  \tablefoot{$^1$~552~AU at the near distance; 836~AU at the far
    distance.  $^2$~500~AU at the near distance; 757~AU at the far
    distance.}
\end{table}

To model this asymmetric behavior, we use a simple parameterized disk
to match the dependence on position angle seen in our MIDI visibility
measurements.  We approximate the disk with a geometrically thin,
optically thick blackbody emitter, with a temperature depending on
radius in the form of a power law in the form of
\begin{equation}
  \label{eq_Tlaw}
  \begin{array}{lr}
    T(r) = T_f \left(\frac{r_f}{r}\right)^{p} & r_i < r \leq r_o,
  \end{array}
\end{equation}
where $T_f$ is the temperature at a fix-point $r_f$, and $r_i$ and
$r_o$ are the inner and outer disk radii, respectively.  We adopt $T_f
= 1500$~K, i.e. the approximate evaporation temperature of silicate
dust, and let $r_f = \sqrt{L/4\pi\sigma{}T_f^4}$, i.e. the radius at
which a blackbody of luminosity $L$ will have a temperature of 1500~K.
For the luminosity of our far model, $r_f = 37$~AU.  We find that
models in which the inner disk radius is at $r_f$ provide poor fits to
our MIDI data, and a substantially smaller inner radius is required.
We adopt $r_i = 1$~AU, which is well below the spatial frequencies
probed by our MIDI measurements.  We thus implicitly assume the inner
disk, consisting of gas, remains optically thick.  This is in line
with interferometric measurements of high-mass young stars, whose disk
emission appears consistently more compact than expected for dusty,
disk-only models \citep[e.g.][]{Monnier05}.

Finally, to account for the more symmetric behavior at shorter
baselines, we include a one-dimensional component in the image
intensity distribution in the form of a Gaussian centered on the disk.
Thus, as fit parameters, we use the outer radius of the disk $r_o$,
the disk inclination $\theta$ and position angle $\phi$, the exponent
$p$ of the temperature power-law, the flux ratio $F_G/F_D$ of the
Gaussian component to the disk component, and the FWHM $\theta_G$ of
the Gaussian component.

\subsubsection{Results}

The parameters that best fit the observed MIDI visibilities were
derived from a brute-force grid search of parameter space, where we
further refine the parameters of the best fit from the grid search by
using the downhill simplex algorithm.  We find that this disk model
produces a very good fit to our data, and in Table~\ref{tab_diskfit},
we summarize the parameters of our best-fit model to the MIDI
visibilities at a wavelength of 10.6~\micron.  We show the resulting
model visibilities as triangles in Fig.~\ref{fig_2d_visfit}, and the
model image in Fig.~\ref{fig_2d_image}.

The observations are best described by a disk spanning up to several
hundred AU, with the disk temperature decreasing gradually as a
function of radius ($T \propto r^{-0.47}$).  The disk is inclined
($\theta=60\degr$), and the semi-major axis is oriented $158\degr$
east of north.  Besides the disk component, a substantial halo is also
present, with a FWHM size equal to about half the diameter of the
disk, and a total flux at $\lambda=10.6$~\micron{} of 1.6 times the
disk component.

\section{Discussion}
\label{sec_discussion}

\subsection{Comparison with previous models}

In order to gauge the importance of including spatial information in
the modeling process, we reconstructed the one-dimensional radiative
transfer model for this object presented by \citet{Guertler91}, using
their dust composition and density structure together with the MoDust
code used for our models.  The model from \citet{Guertler91}, which we
refer to as G91, is a two-zone power-law model, and can be described
using the density prescription from Eq.~\ref{eq_density} if the
parameters $r_1$, $\rho_1$ and $p_1$ are omitted.  In
Table~\ref{tab_modelfit}, we show the parameters for the G91 model
using this prescription.

\begin{figure}[t]
  \begin{center}
    \includegraphics[width=85mm]{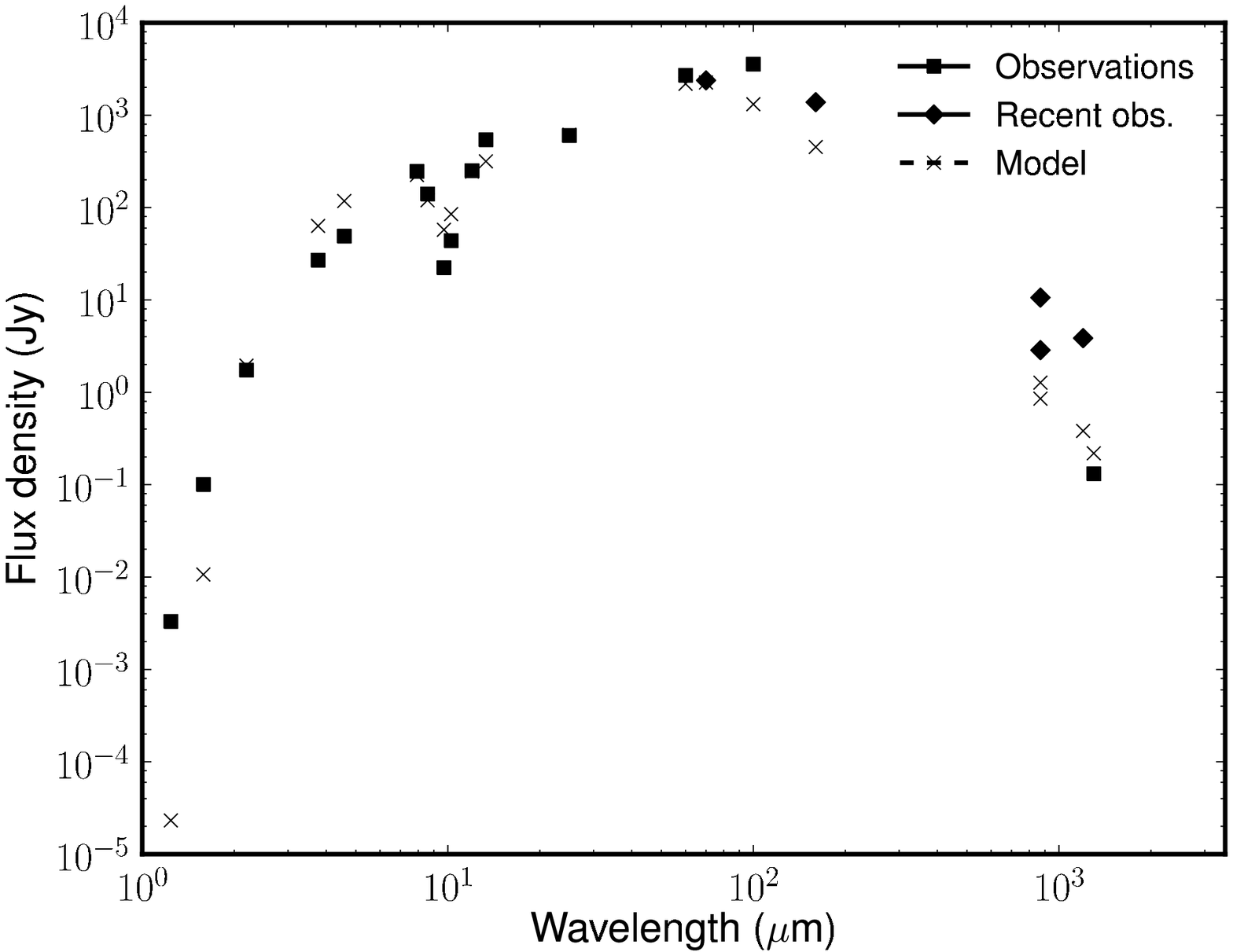}
    \includegraphics[width=85mm]{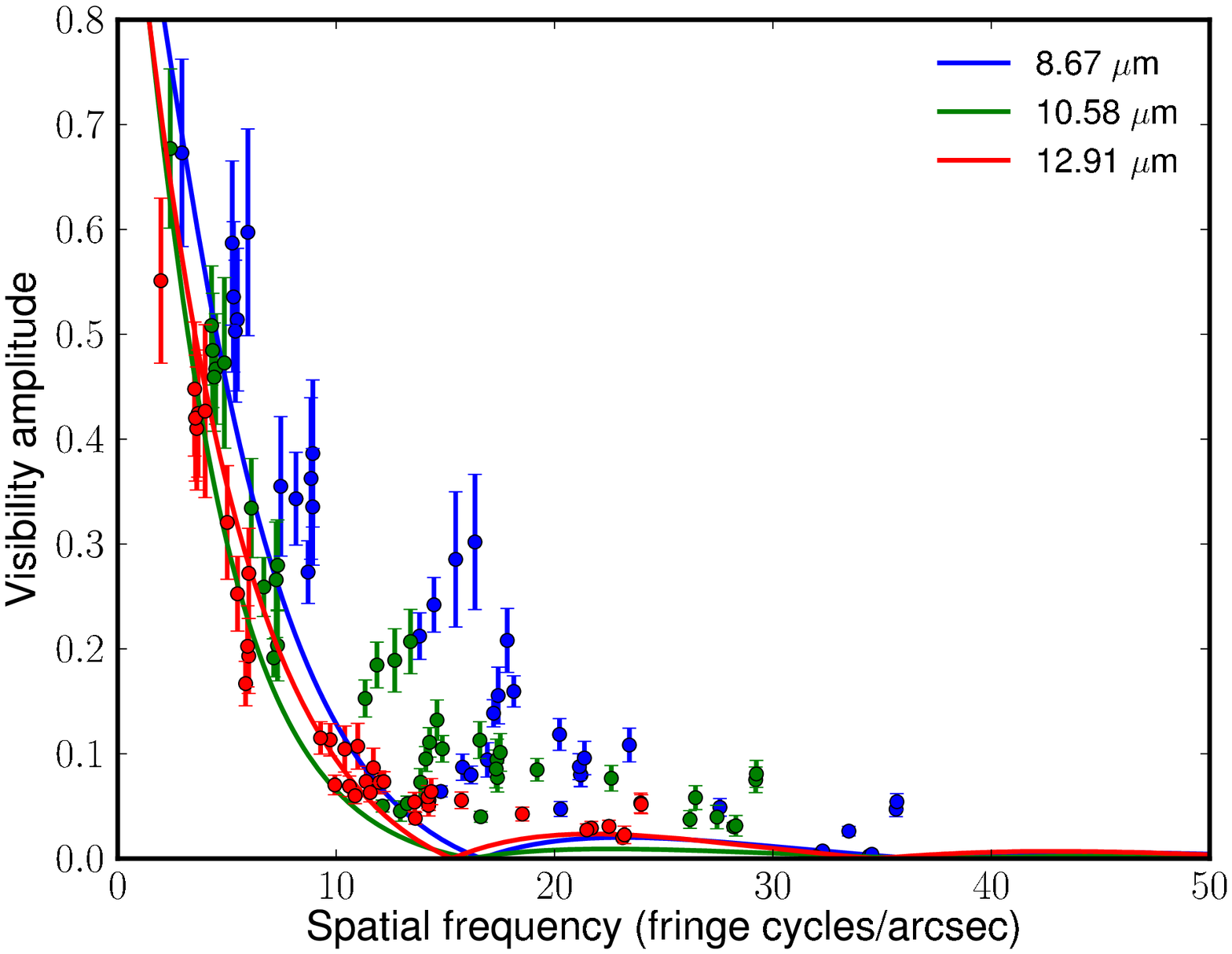}
    \includegraphics[width=85mm]{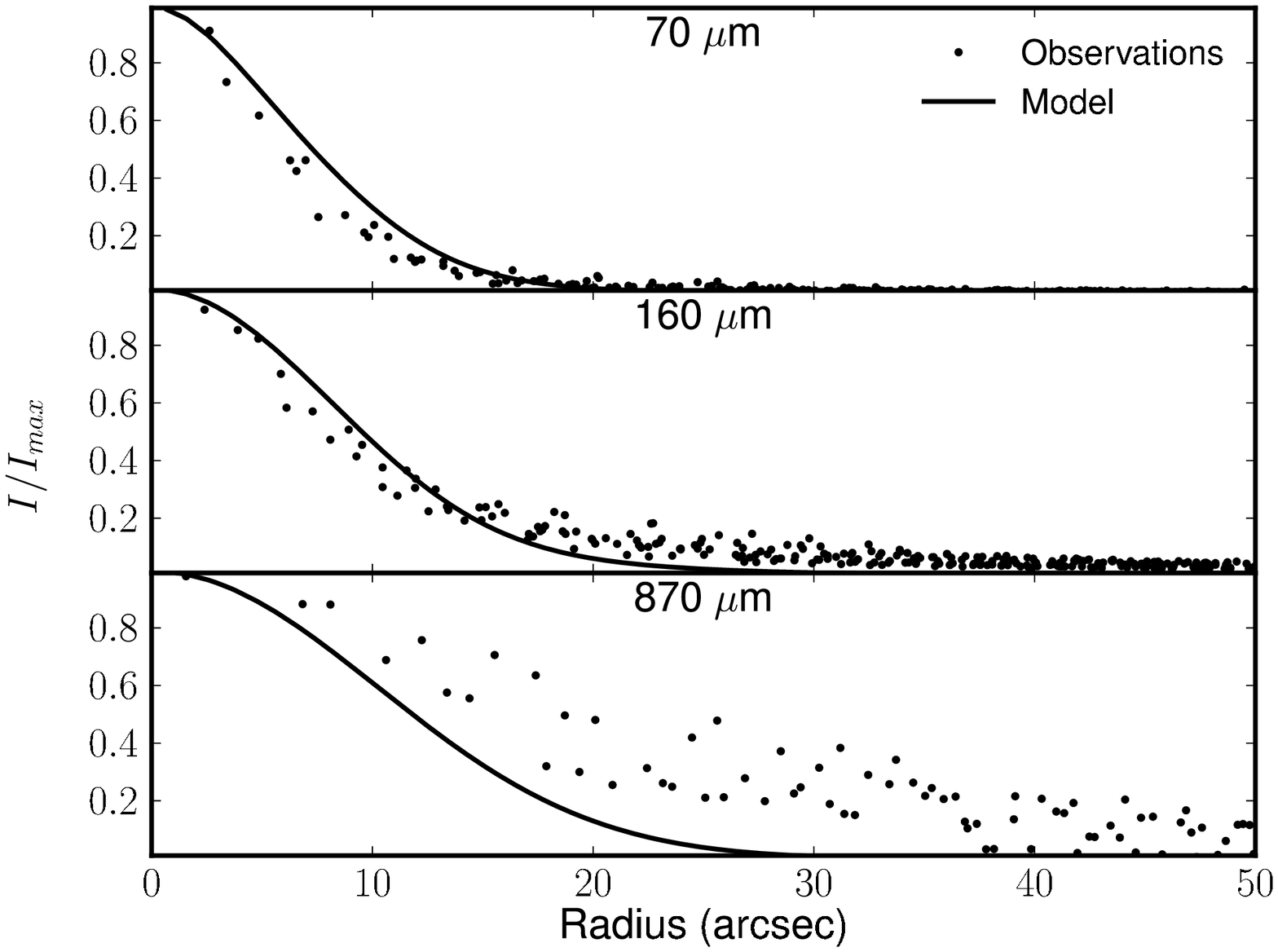}
    \caption{Fit to model by \citet{Guertler91}.  Figure notation is
      the same as for Fig.~\ref{fig_nearfit}, but in the top panel
      of the SED we distinguish between flux values available today
      (filled diamonds) and those available at the time of the
      \citet{Guertler91} study (filled squares).}
    \label{fig_guertler}
  \end{center}
\end{figure}

In Fig.~\ref{fig_guertler} we show the resulting SED, mid-infrared
visibilities, and far-infrared/sub-mm radial intensity profiles.  Note
that \citet{Guertler91} fit only the SED to the extent it was known at
the time, and that they did not have detailed spatial information
(i.e. mid-IR interferometry and far-IR imaging) available.  We
indicate the measurements they had available by squares in
Fig.~\ref{fig_guertler}, while the more recent Hi-GAL and ATLASGAL
measurements, as well as the 1.2~mm measurement from
\citet{Beltran06}, are indicated by diamonds.  Regarding the flux
values available at the time of their study, we note in particular
that their 1.3~mm measurement of 131~mJy was probably in error, as the
position indicated in their Table~2 is more than 50\arcsec{} from the
infrared source.  More recently, for example, \citet{Beltran06}
reported a 1.2~mm flux value of 3.86~Jy for the source, which seems to
be in good agreement with the ATLASGAL flux level of 11.56~Jy at
870~\micron{} which we present here.

\begin{figure}[t]
  \begin{center}
    \includegraphics[width=85mm]{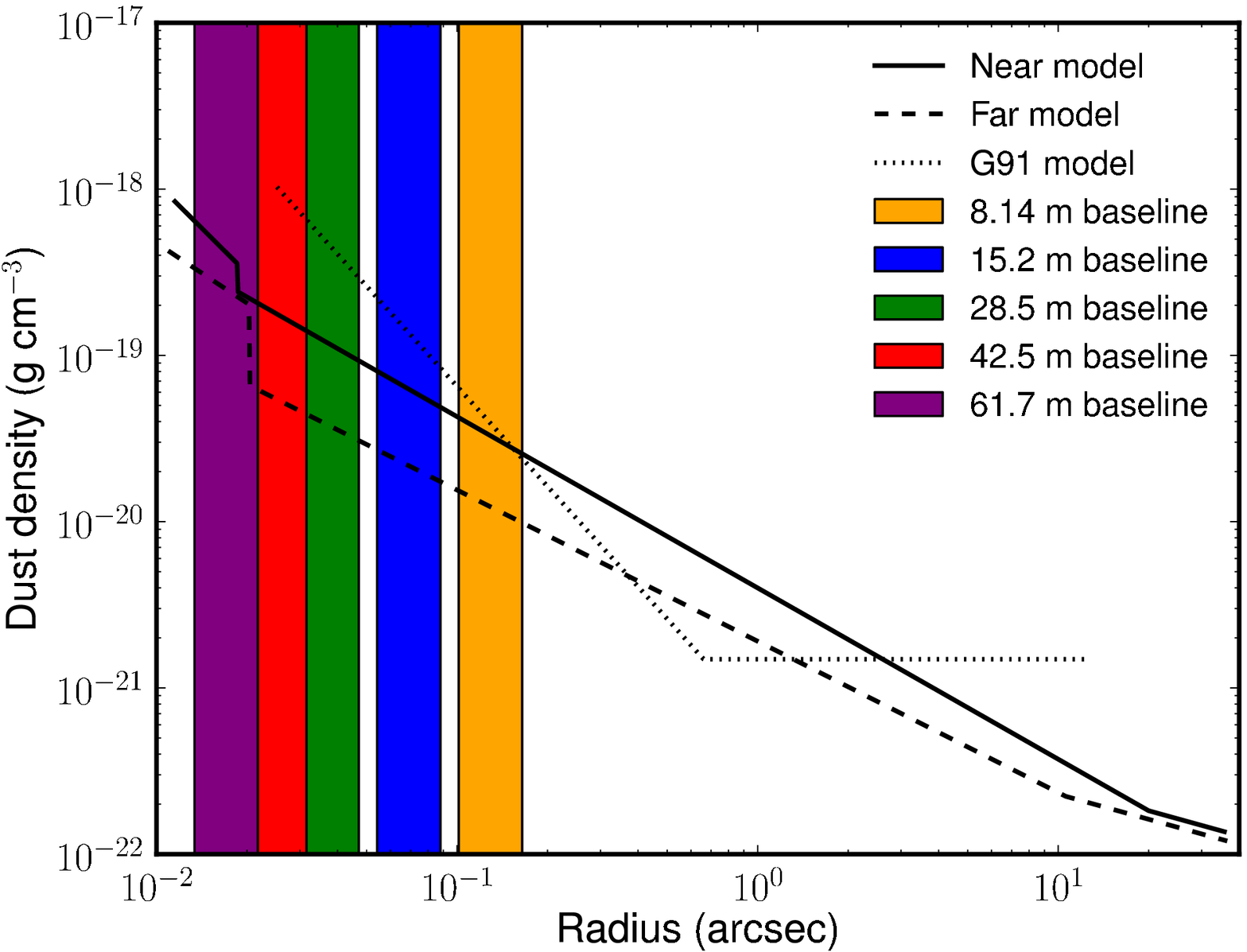}
    \caption{Density laws used for the three radiative transfer
      models, expressed as a function of angular distance on the sky
      to account for the difference in distance to the source each
      model.  The colored rectangles show the spatial scales $\lambda
      / 2B$ covered by the MIDI observations, and the color coding
      matches the average baselines for the groups indicated in
      Fig.~\ref{fig_uvplot}.}
    \label{fig_density}
  \end{center}
\end{figure}

In Fig.~\ref{fig_density}, we show the density laws used for both our
near and far models, as well as the G91 model.  To account for the
different distances to the source used in each model (see
Table~\ref{tab_modelfit}), we express the dust density as a function
of angular position on the sky.  We also overlay the effective angular
scales probed by our MIDI measurements, where have used the same color
coding and grouping by projected baseline as in Fig.~\ref{fig_uvplot}.

For the near and far models, we see that the inner shell with enhanced
density is probed primarily by the longest baseline measurements,
while shorter baselines probe successively larger scales, out to about
0\farcs1.  Because this inner region is completely absent in the G91
model, the mid-IR visibilities at longer baselines are vastly
under-predicted by this model. This is despite the relatively good fit
to the SED (with the exception of the sub-mm/mm regime), as well as
the 70 and 160~\micron{} radial profiles, which only emphasizes the
importance of including spatial information at different scales into
such radiative transfer models.

\subsection{The disk around AFGL~4176}
\label{sec_disc_asym}

The mid-infrared interferometric observations presented in this paper
provide the first evidence for any signs of asymmetry in this
deeply-embedded MYSO to date.  Using a model consisting of a symmetric
halo and an optically thick, geometrically thin disk, we find that a
disk with an inclination angle of $60\degr$ and a position angle of
the semi-major axis of $158\degr$ east of north provides an excellent
fit to our MIDI data.  The best-fit temperature profile of the disk is
$T \propto r^{-0.47}$.  At longer wavelengths, the deviation from
spherical symmetry in our MIDI data is much less pronounced, which
suggests that the emission from the disk is dominated by a more
spherical halo component at these wavelengths.

Nevertheless, because of the wavelength range, mid-infrared
observations probe primarily cooler material ($\sim 200-600$~K),
expected to be well away from the hotter, inner circumstellar regions
where signs of accretion onto the central star may be present.
Furthermore, the inner radius of the disk around AFGL~4176, adopted in
this work to be 1~AU, cannot be determined by our observations, as it
remains unresolved by the MIDI data.  In this context, near-infrared
interferometric observations which probe these inner regions would
make a superb compliment to the current data set.

On the other hand, the lack of detection of an outflow component, or
any other larger-scale asymmetries, remains an open question.  This
may, for example, suggest that the object is still at a very early
stage in the formation process, and that large-scale outflows have not
had time to form.  In any case, far-IR/sub-mm interferometric
observations with ALMA may provide the key to bridging the symmetric
structure on scales of tens to hundreds of thousands of AU to the
asymmetries observed at tens to hundreds of AU.

\subsection{Comparison with other MYSOs}

\begin{table*}
  \begin{center}
    \caption{Summary of parameters for several MYSOs studied on scales
      of $\sim10-100$~AU.  All values approximate.}
    \label{tab_mysos}
    \begin{tabular}{l c c c c c r}
      \hline \hline
      MYSO & $D$ (kpc) & $L$ (\Lsun)& $A_V$ (mag) & \ion{H}{II} region present? & Outflows present? & References \\
      \hline
      AFGL~4176 & $3.5-5.3$ & $10^5$ & $100$ & Yes & ? & 1, 2, 3 \\
      W33A & 3.8 & $10^5$ & 280 & Yes & Yes & 4 \\
      NGC~3603~IRS~9A & 7 & $10^5$ & ? & Yes & ? & 5 \\
      AFGL~2136 & 1.7 & $2-3 \times 10^4$ & 100 & Yes & Yes & 6, 7 \\
      IRAS~13481-6124 & 3.2 & $3 \times 10^4$ & 40? & Yes & Yes & 8 \\
      M8E-IR & 1.5 & $7 \times 10^3$ & 30 & No & Yes & 9 \\
      \hline
    \end{tabular}
  \end{center}
  \tablebib{(1)~This work; (2)~\citet{Phillips98}; (3)~\citet{Ellingsen05}; (4)~\citet{deWit10};
    (5)~\citet{Vehoff10}; (6)~\citet{deWit11}; (7)~\citet{Follert11};
    (8)~\citet{Kraus10}; (9)~\citet{Linz09}.}
\end{table*}

To date, very few MYSOs have been studied on scales of tens to
hundreds of AUs.  Here we compare our results with similar studies of
five other MYSOs.  In Table~\ref{tab_mysos} we list the luminosities,
the depth of the 10~\micron{} silicate absorption feature, whether or
not any compact \ion{H}{II} region is present, and whether or not any
outflows have been detected, which can be useful for independently
constraining disk geometry.  As previously noted, however, in the case
of AFGL~4176, no outflow components have been detected
\citep{deBuizer03,deBuizer09}.

The sources W33A and NGC~3603~IRS~9A have similar luminosities to
AFGL~4176, on the order of $10^5~L_\odot$, and both have associated
hyper-compact \ion{H}{II} regions.  However, IRS~9A seems to be much
less embedded: besides having an optical component, the $N$-band
spectrum shows no 10~\micron{} silicate feature at all.  At
mid-infrared wavelengths, the source is very extended, and completely
overresolved in MIDI observations probing $\sim200$~AU scales
\citep{Vehoff10}.  W33A, on the other hand, could be even more
embedded than AFGL~4176 \citep[$A_V \ga 200$~mag,][]{deWit10}, but the
$N$-band visibility levels are poorly determined due to the low amount
of (correlated) flux in the deep 10~\micron{} silicate absorption
feature.  The source was modeled by \citet{deWit10}, who concluded
that most of the mid-infrared emission on scales of $10^2$~AU is
dominated by collimated outflow cones, without compelling evidence for
significant emission from a dusty disk.

At an estimated luminosity of $\sim 5 \times 10^4$~\Lsun, the SED of
AFGL~2136 closely resembles that of AFGL~4176, including a similar
depth of the 10~\micron{} silicate feature \citep{Smith00}.  Using
visibility measurements obtained with MIDI, both \citet{deWit11} and
\citet{Follert11} reported mid-IR emission scales of $\sim 200$~AU,
while \citet{Follert11} found an elongated structure at these scales,
parallel to earlier predictions of disk orientation.

Despite having a lower luminosity, on the order of
$3\times10^4~L_\odot$, the SED of IRAS13481-6124 is somewhat similar
to that of AFGL~4176, although the 10~\micron{} silicate feature is
much shallower in IRAS13481-6124.  \citet{Kraus10} discovered a
molecular outflow, and used near-IR interferometric observations to
resolve an elongated structure oriented perpendicular to the outflow
direction.  The extended outflow shocks detected at near-IR
wavelengths at large distances from the source (Stecklum et
al. 2010\footnote{http://www.jcu.edu.au/hmsf10/Presentations/Monday/Session3/Stecklum.pdf}),
as well as the radiative transfer models presented by \citet{Kraus10},
suggest the disk is strongly inclined.  However, the 20~\Msun{} (gas +
dust mass) disk at scales of 5-150~AU proposed by \citet{Kraus10} is
clearly a fundamentally different model than the structures we present
here for AFGL~4176.  For comparison, the dust mass enclosed in the
near and far models for $r < 150$~AU is $8\times10^{-6}$~\Msun{} and
$4\times10^{-5}$~\Msun{}, respectively, although a direct comparison
between studies using near- and mid-IR interferometric measurements is
difficult.

Finally, the source with the lowest luminosity, M8E-IR, has an overall
extinction of $A_V \approx 30$~mag, as determined from the depth of
the 10~\micron{} silicate feature \citep{Linz09}.  Outflow signatures
have been detected for this source \citep{Mitchell88}.  No cm
continuum emission has been detected directly coinciding with the IR
object.  This lack of an ultra-compact \ion{H}{II} region might be
related to the hypothesis of a bloated, cooler central object, as
suggested by \citet{Linz09}.

Given the small selection of sources for which this information is
available, it is not possible to infer clear, empirical correlations
between the properties listed in Table~\ref{tab_mysos}.  Of the
sources presented, none of them seem to be direct analogues of
AFGL~4176, however W33A might be the closest match in terms of
luminosity, SED and emission scales at mid-infrared wavelengths.
Suggestions by \citet{deWit11} and \citet{Linz11} indicate that mid-IR
interferometric observations of at least some MYSOs may be heavily
influenced by outflow components.  On the other hand, the mid-IR
visibilities presented by \citet{Follert11} apparently trace the disk
structure of AFGL~2136.  Taken together, it becomes clear that the
mid-IR emission structure of MYSOs at scales of tens to hundreds of
AUs can be complex, and caution should be exercised in interpreting
data sets with limited $uv$ coverage.

\section{Summary and conclusions}

In this paper, we reported on spatially resolved observations and
modeling of the massive young stellar object AFGL~4176.  We presented
the largest amount of $N$-band visibilities for any MYSO to date,
obtained with MIDI on the ESO/VLTI, together with maps of extended
emission at 70 and 160~\micron{} from the Hi-GAL survey on the
Herschel Space Observatory and 870~\micron{} imaging from the ATLASGAL
survey with the APEX telescope.

With the exception of the mid-infrared interferometric measurements
presented in this paper, the observational data available for this
deeply-embedded object do not provide indications of asymmetry or
preferred geometry.  In view of this, we used spherically symmetric
radiative transfer models, consisting of an envelope with a density
enhancement at its inner edge, to model the source.  Using these
one-dimensional models, we were able to satisfactorily reproduce,
simultaneously, the SED and spatial structure at mid-IR through mm
wavelengths.

However, the one-dimensional models by definition cannot reproduce the
asymmetries present in the MIDI data.  We therefore modeled these data
separately, using a multiple-component geometric analysis.  We
interpret the MIDI observations in terms of a circumstellar disk
around AFGL~4176.  If this disk hypothesis is correct, it would make
AFGL~4176 the most luminous (massive) young star to have a disk
detected around it to date.  However, beyond initial detection, much
remains to be clarified regarding the nature and extent of the disk,
making this object a top priority for additional interferometric
imaging, at both near-IR and far-IR/sub-mm wavelengths.

\newpage

\begin{acknowledgements}
We would like to thank our referee, Dr. Koji Murakawa, for insightful
critique during the review process of this manuscript.  We thank
Christoph Leinert, Svitlana Zhukovska, Dima Semenov and Keiichi Ohnaka
for useful discussions.  We also express our gratitude to Frederic
Schuller for generously making the reduced ATLASGAL observations
available to us, and \c{S}eyma \c{C}ali\c{s}kan for supporting work
during an earlier phase of this project.  AMS was partly supported by
Russian state contract No. 16.518.11.7074 and the Russian Foundation
for Basic Research (grants 11-02-01332 and 11-02-97124-p).
\end{acknowledgements}

\bibliography{refs}

\end{document}